\documentclass[twocolumn]{aastex63}
\usepackage{graphics,graphicx}
\usepackage{amsmath}

\definecolor{webgreen}{rgb}{0,.5,0}
\definecolor{webbrown}{rgb}{.6,0,0}
\definecolor{pink}{rgb}{0.858, 0.188, 0.478}

\newcommand{\kmps}{km s$^{-1}$}
\definecolor{darkgreen}{rgb}{0.13, 0.55, 0.13}

\newcommand\hi{\mbox{\ion{H}{1}}}

\newcommand\pcc{\;{\rm cm}^{-3}}
\newcommand\Msun{\; M_{\odot}}
\newcommand\kms{\; {\rm km}\;{\rm s}^{-1}}

\newcommand\erg{\; {\rm erg}}

\newcommand\yr{\; {\rm yr}}

\newcommand\Myr{\;{\rm Myr}}

\newcommand\pc{\;{\rm pc}}
\newcommand\kpc{\;{\rm kpc}}

\newcommand\sfrunit{\Msun \kpc^{-2} \yr^{-1}}

\newcommand\Kel{\;{\rm K}}

\newcommand\simgt{\lower.5ex\hbox{$\; \buildrel > \over \sim \;$}}
\newcommand\simlt{\lower.5ex\hbox{$\; \buildrel < \over \sim \;$}}
\newcommand\pderiv[2]{\frac{\partial {#1}}{\partial {#2}}}
\newcommand\deriv[2]{\frac{d {#1}}{d {#2}}}

\newcommand\rbrackets[1]{\left({#1}\right)}
\newcommand\sbrackets[1]{\left[{#1}\right]}

\newcommand\abrackets[1]{\left\langle{#1}\right\rangle}
\newcommand\divergence[2][\rbrackets]{\nabla \cdot #1{#2}}

\newcommand\vel{\mathbf{v}}

\newcommand\xhat{\hat{\mathbf{x}} }
\newcommand\yhat{\hat{\mathbf{y}} }
\newcommand\zhat{\hat{\mathbf{z}} }

\newcommand\Fmom{\mathcal{F}_p}

\newcommand\Fph[2]{\mathcal{F}_{{#1},{\rm {#2}}}}

\graphicspath{{figures/}}

\shorttitle{Multiphase Galactic Outflows}
\shortauthors{Vijayan et al.}

\begin{document}

\title{Kinematics and Dynamics of Multiphase Outflows in
Simulations of the Star-Forming Galactic ISM
}
\email{aditiv@rri.res.in,cgkim@astro.princeton.edu}
\email{lucia.armillotta@anu.edu.au,eco@astro.princeton.edu}
\email{mli@flatironinstitute.org}

\author[0000-0002-7714-2379]{Aditi Vijayan}
\affil{Raman Research Institute, Bangalore, 500080, India}
\affil{Joint Astronomy Programme, Department of Physics, Indian Institute of Science, Bangalore 560012, India}

\author[0000-0003-2896-3725]{Chang-Goo Kim}
\affiliation{Department of Astrophysical Sciences, Princeton University, Princeton, NJ 08544, USA}
\affiliation{Center for Computational Astrophysics, Flatiron Institute, New York, NY 10010, USA}

\author[0000-0002-5708-1927]{Lucia Armillotta}
\affiliation{Research School of Astronomy and Astrophysics, The Australian National University, Canberra, ACT, 2611, Australia}

\author[0000-0002-0509-9113]{Eve C. Ostriker}
\affiliation{Department of Astrophysical Sciences, Princeton University, Princeton, NJ 08544, USA}

\author[0000-0003-0773-582X]{Miao Li}
\affiliation{Center for Computational Astrophysics, Flatiron Institute, New York, NY 10010, USA}

\begin{abstract}

Galactic outflows produced by stellar feedback are known to be multiphase in nature. Both observations and simulations indicate that the material within several kpc of galactic disk midplanes consists of warm clouds embedded within a hot wind.  A theoretical understanding of the outflow phenomenon, including  both winds and fountain flows, requires study of the interactions among thermal phases. We develop a method to quantify these interactions via measurements  of mass, momentum, and energy flux exchanges 
using temporally and spatially averaged quantities and conservation laws. 
We apply this method to a star-forming ISM MHD simulation based on the TIGRESS framework, for Solar neighborhood conditions.
To evaluate the extent of interactions among the phases, we first examine the validity of the ``ballistic model,'' which predicts 
trajectories of the warm phase ($5050\,\rm{K}<T<2\times10^4\,\rm{K}$) treated as non-interacting clouds. 
This model is successful at 
intermediate vertical velocities ($ 50 \kms\lesssim |v_z| \lesssim 100 \kms$),
but at higher velocities we observe an excess in simulated warm outflow compared to the ballistic model.
This discrepancy cannot be fully accounted for by cooling of 
high-velocity intermediate-temperature  ($2\times10^4\,\rm{K}<T<5\times10^5\,\rm{K}$) gas. By examining the fluxes of mass, momentum and energy, we conclude that warm phase gains mass via cooling of the intermediate phase, while momentum transfer occurs from the hot ($T>5\times10^5\,\rm{K}$)  to the warm phase.
The large energy flux from the hot outflow that is 
transferred to the warm and intermediate phases is quickly radiated away. 
A simple interaction model implies an effective warm cloud size in the fountain flow 
of a few 100~pc,  showing that warm-hot flux exchange 
mainly involves a few large clouds rather than many small ones.

\end{abstract}


\section{Introduction} \label{sec:intro}

The  formation and evolution of galaxies are regulated by 
accretion  and expulsion of gas.  
Thus, characterizing the nature and evolution of galactic outflows is of fundamental importance to understanding galaxy evolution. For galaxies that are less massive than the Milky Way, the main mechanism believed to be responsible for ejecting
interstellar medium (ISM) 
is stellar feedback \citep[e.g.,][]{Somerville15, Naab17}. Supernova-driven galactic outflows indeed have been directly observed in several nearby star-forming galaxies, revealing a complex multiphase nature: they are composed of hot (T~$\sim 10^{6-8}$~K) \citep[e.g.,][]{Strickland+04, Strickland&Heckman07}, warm ionized (T~$\sim 10^{5}$~K), neutral (T~$\sim 10^{4}$~K) \citep[e.g.,][]{Martin05,Teng+13,Chen+10,Heckman+15,Chisholm+17}, and cold (T~$\sim 10^{1-3}$~K) molecular \citep[e.g.,][]{Walter+02, Bolatto+13, Leroy+15} and atomic gas \citep[e.g.,][]{Martini+18}.

In the Milky Way, presence of gas in extra-planar regions  ($\sim 1-2$~kpc above the disk) has been mainly probed by neutral \hi\ \citep[the ``Lockman layer,'' e.g.,][]{Lockman84, Lockman02, Ford+10, Peek+11} and ionized H$\alpha$ emission \citep[the ``Reynolds layer,'' e.g.,][]{Reynolds91,Haffner+03,Gaensler+08}. This extra-planar gas is characterized by disk-like kinematics and metallicities close to the solar value, suggesting a galactic origin \citep[e.g.][]{vanWoerden+04, Marasco&Fraternali11, Qu&Bregman19}. Moreover, models aiming to reproduce the extra-planar gas kinematics have shown a negative vertical gradient of the rotational velocity (often called a ``lag'') and a global infall motion \citep{Marasco&Fraternali11}, lending support to a fountain flow origin \citep[e.g.,][]{Shapiro&Field76, Bregman80}.
Similar kinematic signatures are now commonly observed in nearby edge-on galaxies from sensitive \hi\ observations \citep[e.g.,][]{Fraternali+02, Zschaechner+11, Marasco+19} and recent integral field unit surveys of extra-planar H$\alpha$ \citep[e.g.][]{Bizyaev+17, Levy+19}.

Cosmological simulations \citep[e.g.,][]{Schaye+15,Pillepich+18} of galaxy formation focus on the large scale effects of feedback in the redistribution of gas within and outside galaxies, and the gas flow to and from the circum-galactic medium (CGM). 
However, large-scale simulations lack sufficient resolution 
to follow the gas dynamics driven by feedback explicitly. Instead,  phenomenological models for feedback (combined with subgrid models of ISM and star formation) are adopted and tuned to yield consistency with the stellar mass-halo mass relation at redshift zero \citep[e.g.,][and references therein]{Somerville15, Naab17}. To improve the predictive power of large-scale simulations, an important next step is to replace phenomenological subgrid models with models in which the mass (hydrogen and metal), momentum, and energy fluxes in outflows are calibrated from simulations in which feedback effects and multiphase gas are resolved throughout the volume. 

To this end, wind mass loading factors (mass outflow rate normalized by star formation rate) have been measured from isolated galaxy simulations and cosmological zoom simulations \citep[e.g.,][]{Hopkins12,Muratov+15,Christensen+16,Angles-Alcazar+17,Marinacci+19A}. Still, such simulations cannot achieve high enough resolution to resolve the Sedov-Taylor stage of individual SNR evolution.  Instead, they employ momentum injection schemes  \citep[e.g.,][]{Kimm+14,KO15a,Hopkins18}, which do not allow for creation of hot gas from blastwaves \citep[e.g.,][]{Rosdahl+17,Smith+18,Hu19} except possibly for dwarf galaxies or sufficiently high resolution Milky-Way models. Furthermore, as resolution gets poorer outside of galactic disks, different thermal gas phases in outflows often cannot be resolved. As a result, 
the reported mass loading factors in zoom and global galaxy simulations are effectively for a numerically-mixed, single phase outflow.
This approach does not properly account for stark differences that may exist between the final fate of cool and hot outflowing phases.  
In  particular, the cooler phase may have insufficient momentum flux to escape the gravitational potential of a galaxy (except dwarf galaxies) and would form a fountain flow,
while the hot material can easily escape as a wind.  
While simulations that do not adequately resolve extra-planar regions may suffer from overmixing,  interactions among phases may nevertheless be quite important, and may alter the evolution from what would occur in isolation.     

Physical processes driving  the  interaction  of  different  gas  phases in extra-planar regions  have commonly been investigated via small-scale idealized simulations  \citep[e.g.][and references therein]{Scannapieco&Bruggen15, Schneider17, Gronke+18, Sparre19}.  These ``shock-cloud interaction'' simulations model the evolution of warm gas clouds, representing disk material ejected by stellar feedback, interacting with a more tenuous medium, representing the hotter, higher-velocity outflow phase. These idealized simulations focus on questions such as whether cool clouds can be accelerated without being destroyed, and whether interactions can induce cooling of hot gas. However, a complete characterization of extra-planar gas flows requires analysis of simulations that 
follow both the origin and evolution of outflowing material, so that space-time relationships between hot winds and cooler embedded structures will be realistic. This necessitates a realistic treatment of the multiphase ISM where outflows originate, including self-consistently regulated  star formation and stellar feedback.

Recently, ``local patch'' simulations of galactic disk regions have begun to incorporate self-consistent star formation (through gravitational collapse) and supernova feedback \citep[e.g.][]{Gatto+17,Iffrig&Hennebelle17,Kim&Ostriker17} that bridge the gap between large- and small-scale simulations. One of the chief advantages of local disk simulations is that the uniformly high resolution affords a thorough investigation of multiphase outflows \citep[e.g.,][]{Kim&Ostriker18}. 
Well-resolved multiphase outflows, driven by explicitly modelled physical mechanisms, allow us to investigate not only the out-going mass, momentum, and energy budgets from different phases separately \citep[e.g.,][]{Li+17,Fielding+18}, but also detailed kinematics and dynamics of gas flows and interaction between thermal phases.
\citet{Kim&Ostriker18} emphasize the importance of spatial resolution in extra-planar regions of both cool and hot phases.  
Since mass and energy fluxes are respectively dominated by warm fountains and a hot wind, a numerically-mixed outflow might result in 
incorrect mass, metal, and/or energy outflow rates.

In this paper, we introduce a method to investigate multiphase galactic outflows (wind and fountain flows), testing the conservation and exchange of mass, momentum, and energy between different thermal phases. We apply this analysis to the  solar  neighborhood  model simulated using the TIGRESS (Three-phase ISM in Galaxies Resolving Evolution with Star formation and Supernova feedback) framework 
for 3D MHD models of the star-forming ISM in galactic disks \citep{Kim&Ostriker17}. We first characterize the kinematic and dynamical properties of different gas phases. We then compare these properties with the predictions of a simple ballistic model.  Finally, we use the  conservation laws of hydrodynamics to reveal mass, momentum, and energy flux exchanges between phases.

The paper is organized as follows: In Section~\ref{sec:tigress}, we briefly introduce the TIGRESS framework and discuss the overall extra-planar gas distribution, the time evolution of outflows, and the horizontally and temporally averaged outflow properties in the simulation. In Section \ref{sec:ballistic}, we compare the time-averaged velocity probability distribution functions from the simulation with predictions of a model in which individual fluid elements follow ballistic trajectories. In Section \ref{sec:flux_analysis}, we analyze mass, momentum, and energy fluxes to understand how different thermal phases interact with each other during the outflow evolution. We also introduce a simple interaction model to estimate the  effective size of clouds.
We conclude with a summary of our results and discussion in Section~\ref{sec:summary}.

\section{Solar Neighborhood TIGRESS Model}\label{sec:tigress}

\subsection{TIGRESS framework}\label{subsec:tigress_summary}
The simulation analyzed in this work was performed using the TIGRESS framework \citep{Kim&Ostriker17}, in which the star-forming ISM is self-consistently modelled. Here we use results from the solar neighborhood model, whose outflow properties are presented in \citet{Kim&Ostriker18}. In the TIGRESS framework, the ideal magnetohydrodynamics (MHD) equations are solved using the Athena code \citep{Stone+09,Stone+08} in a local box
representing a small patch of differentially rotating galactic disk. The solar neighborhood model adopts a  galactic rotation rate  $\Omega(R_0)=28\kms{\,\rm kpc^{-1}}$ for the center of the domain, with a flat rotation curve $q\equiv -d\ln\Omega/d\ln R = 1$ that gives rise to ordered shear of the background azimuthal velocity  (local $\yhat$ direction) 
varying along the local radial direction ($\xhat$). The horizontal extent of the simulation domain is -512~pc~$<x,y<$512~pc, with periodic boundary conditions in $y$ and shearing-periodic boundary conditions in $x$ \citep{Stone+10}. The vertical domain extends to $z=\pm$3584~pc and has outflow boundary conditions. The entire numerical domain has uniform resolution of 4~pc. The simulation includes gas self-gravity, optically thin cooling, and spatially uniform grain photoelectric heating by far-ultraviolet (FUV) radiation. Sink particles are employed to trace creation of and gas accretion onto star clusters in regions with unresolved collapse. Gravity from and on cluster particles is computed using particle-mesh methods. The stellar feedback from star clusters is modelled in the form of supernova explosions and FUV radiation based on the {\tt STARBURST99} population synthesis model \citep{Leitherer+99}. We refer readers to \citet{Kim&Ostriker17} for more details.

External gravity from the old stellar disk and the dark matter halo
is modelled with a fixed gravitational potential that varies only in the $\zhat$ direction \citep{Kujiken&Gilmore}. The functional form used in the simulation and our analysis is  
\begin{equation}
\begin{split}
\Phi_{\mathrm{ext}}(z) = \; & 2 \pi G \Sigma_*z_* \sbrackets{ \rbrackets{ 1 + \frac{z^2}{z_*^2}}^{1/2} -1}\\
&+ 2 \pi G \rho_\mathrm{dm} R_0^2 \,\ln\rbrackets{ 1 + \frac{z^2}{R_0^2}}\,,
\end{split}
\label{eqn:pot}
\end{equation}
where $\Sigma_* = 42$ M$_{\odot}$ pc$^{-2}$, $z_*=245$ pc, $\rho_{\rm{dm}} = 0.0064$ M$_{\odot}$ pc$^{-3}$ and $R_0 = 8$ kpc.

\subsection{Extra-planar Gas Distribution}\label{subsec:tigress_gas_dist}

The simulation begins with a vertically-stratified, horizontally uniform gas distribution. The initial conditions assume a double exponential density profile, which soon cools (creating a cold phase), with some of the gas collapsing into dense clouds to form star clusters. Young massive stars in newly formed star clusters produce FUV radiation that heats the warm and cold medium, and eventually explode as SNe that create hot gas and drive turbulence. After the first star formation burst and feedback cycle, the system adjusts its global star formation rates to a self-regulated state in which the feedback maintains the turbulent, thermal, and magnetic support needed to offset the vertical weight of the gas \citep[e.g.,][]{KOK13, KO15b, Kim&Ostriker17}.

\begin{figure} 
	\centering
	\includegraphics[width=\columnwidth]{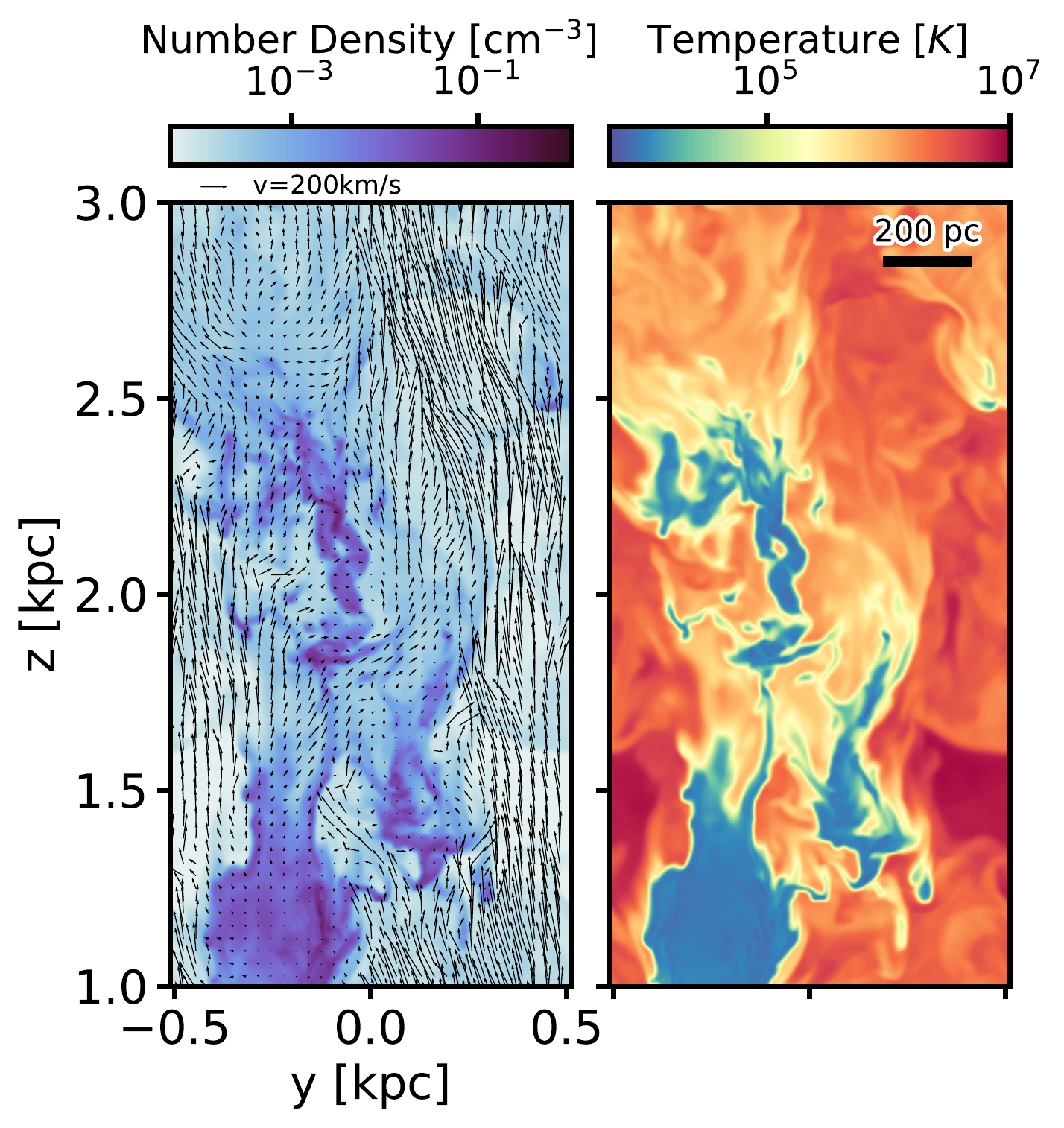}
	\caption{Sample slices at $t=440\Myr$ through $x=140\pc$ showing number density (left panel) and  temperature (right panel) of the gas for $z>1\kpc$. The arrows overlapping the density slice represent the gas velocity field. Arrow lengths indicate velocity magnitudes. The multiphase nature of the outflowing gas is clearly visible in the temperature slice. Slow-moving warm and dense clouds are surrounded by fast-moving hot and rarefied gas, with intermediate temperature gas at interfaces and in  wakes.} 
	\label{fig:slicenT}
\end{figure}

While a self-regulated equilibrium state is achieved within the gas disk near the midplane (within the gas scale height, $H\sim 300-400\pc$), some of the gas heated and accelerated by SN blastwaves breaks out into the extra-planar region ($|z|>(1-2)H$). The outflowing gas includes not only the hot, shock-heated component, but also the highest-velocity portion of the warm gas accelerated by superbubble expansion 
(\citealt{Kim&Ostriker18}; see also \citealt{KOR2017}). To visualize the multiphase structure of gas in the extra-planar region, in Figure~\ref{fig:slicenT} we show $\yhat$-$\zhat$ slices of gas number density overlaid with velocity field (left),  and the temperature (right). We select a simulation snapshot at $t=440\Myr$ when there is a strong outflow (see below) 
and slice through $x=140\pc$. The multiphase nature of the outflowing gas is clearly visible. Warm ($T\sim10^4\Kel$) and dense ($n\sim0.1\pcc$) clouds moving with relatively low velocity (few $10$s \kmps) are surrounded by tenuous ($n \lesssim 10^{-3}\pcc$), hot ($T>10^{6-7}\Kel$) gas with $v_z>200\kms$. The intermediate density and temperature gas is visible at the interface between warm and hot gas and behind the warm clouds as wakes.

\begin{figure} 
	\centering
	\includegraphics[width=\columnwidth]{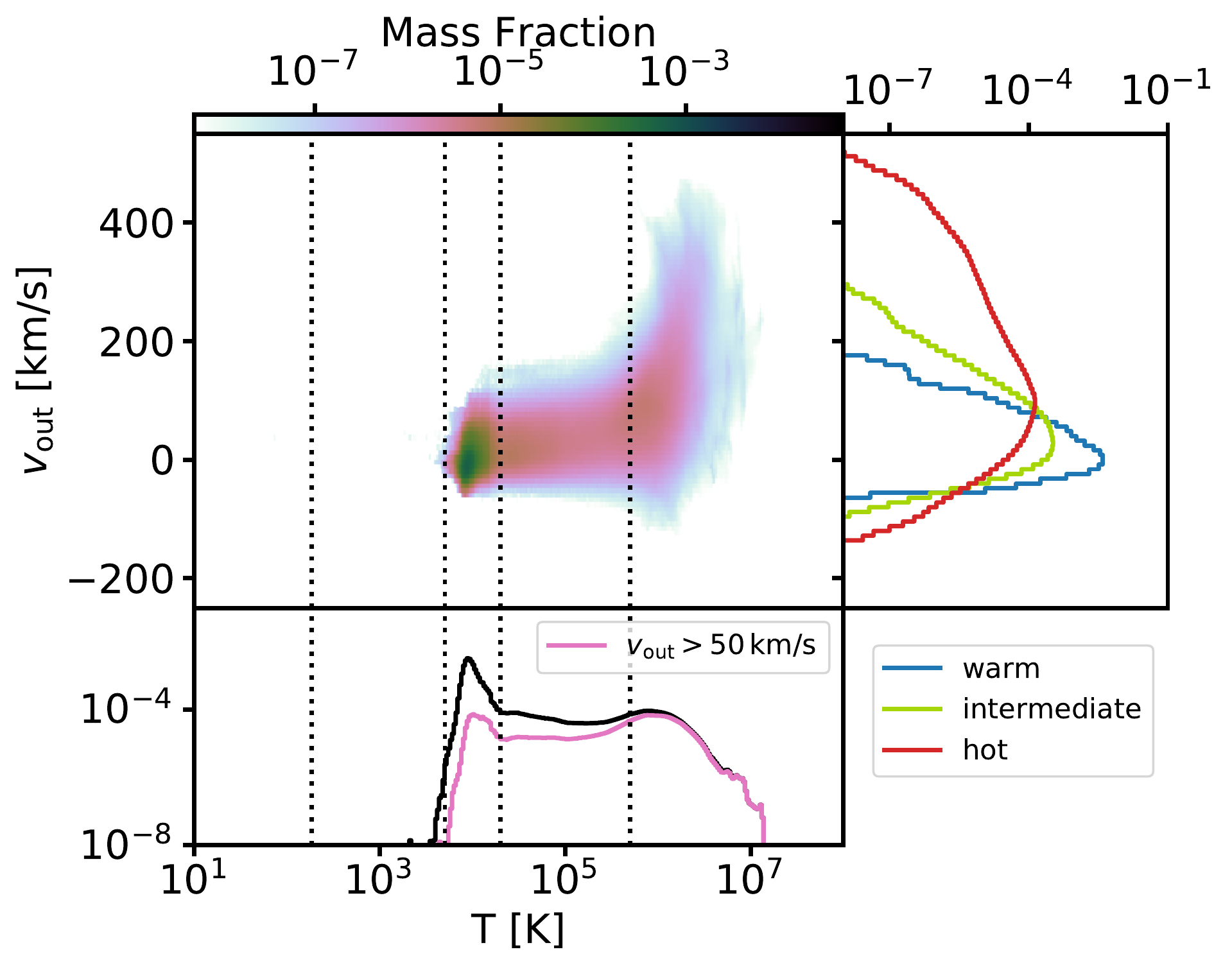}
	\caption{Mass-weighted joint PDF of the extra-planar gas ($z>1\kpc$) from the snapshot at $t=440\Myr$ in the temperature and outflow velocity ($v_{\rm out} \equiv v_z{\rm sign}(z)$) plane. The vertical dotted lines demark of the thermal phases. 
	The bottom panel shows the temperature PDF (velocity-integrated, black line; high velocity, pink line). The right panel shows the phase-separated velocity PDFs (key  lower right).}
	\label{fig:pdfs}
\end{figure}

For a more quantitative view of the gas distribution, Figure~\ref{fig:pdfs} shows a mass-weighted joint probability distribution function (PDF) of all the extra-planar gas (${z}>1\kpc$) for the same time as in Figure~\ref{fig:slicenT}. The mass within a given temperature and outflow velocity ($v_{\rm out} \equiv v_z{\rm sign}(z)$) bin is normalized by the total gas mass in the simulation. Note that the extra-planar region comprises only $3\%$ of the total gas mass.
We demark the thermal phases using vertical dotted lines, and refer to the gas with $5\times10^3~\rm{K}<\rm{T}<2\times10^4~\rm{K}$, $2\times10^4~\rm{K}<\rm{T}<5\times10^5~\rm{K}$, and $5\times10^5\Kel<T$ as warm, intermediate, and hot, respectively (the same demarcations as in \citealt{Kim&Ostriker17}). In the bottom panel, the temperature histogram of all the gas (black curve) shows the dominance of the warm medium by mass ($>90\%$). However, if we only consider ``high-velocity'' gas ($v_{\rm out}>50\kms$, pink curve), the warm and hot components have comparable mass. In the right panel, the phase-separated velocity histograms\footnote{We adopt a consistent color scheme throughout the paper to distinguish thermal phases: blue for warm, lime green for intermediate, and red for hot.} show clear development of outflow tails (positive $v_{\rm out}$). We analyze the velocity PDF in greater detail in Section \ref{sec:ballistic}. 

\subsection{Time Evolution}\label{subsec:tigress_time_evol}

\begin{figure*}
    \centering
    \includegraphics[width=\textwidth]{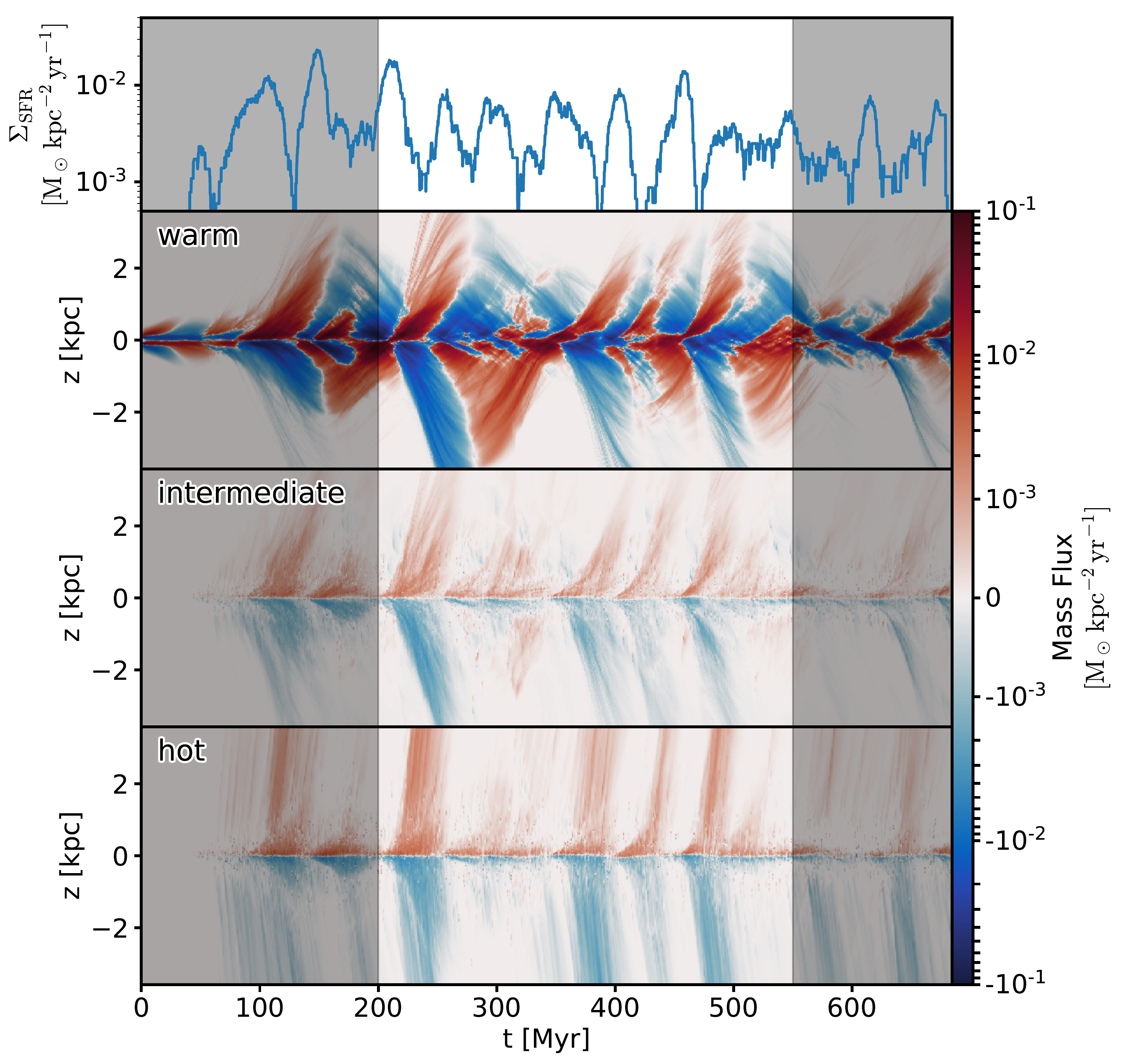}
    \caption{Space-time diagram of mass fluxes for warm, intermediate, and hot phases (bottom three panels) along with the SFR surface density (top panel). Detailed analyses of the simulation are for the central $t=200-550\Myr$ (gray shaded regions are not included). Star formation is bursty, with the period of $40-50\Myr$ corresponding to the vertical crossing time scale of the diffuse warm/cold medium. 
    For the portion analyzed, the 16th, 50th, and 84th percentiles of the SFR surface density are $\Sigma_{\rm SFR} = 2.8^{+4.2}_{-1.5}\times10^{-3}\sfrunit$.
    }
    \label{fig:mflux}
\end{figure*}

We construct horizontally-averaged profiles to understand the overall time evolution of gas flows. For each thermal phase, the horizontal average of a quantity $q$ is defined by
\begin{equation}\label{eqn:havg}
\overline{q}_{\rm ph}(z;t) = \frac{\sum_{x,y} q(x,y,z;t)\Theta_{\rm ph}(T) \Delta x \Delta y}{L_x L_y},
\end{equation}
where $\Theta_{\rm ph}(T)$ is the top-hat function that returns 1 for gas at temperatures within the temperature range of each phase (ph = warm, intermediate, or hot) or 0 otherwise, $\Delta x = \Delta y =4\pc$ is the cell size, and $L_x=L_y=1024\pc$ is the horizontal domain size.

Figure~\ref{fig:mflux} shows the time evolution of the horizontally-averaged mass fluxes, $\overline{\rho v_z}$, for the warm, intermediate, and hot phases. The red/blue color denotes positive/negative mass flux. With this convention, red/blue corresponds to outflow/inflow for the gas above the midplane, $z>0$; the opposite is true for the lower half of the disk, $z<0$. The time evolution of the warm mass flux at a given height alternates color, evidencing a fountain in which the majority of the warm outflow falls back since it has been launched with insufficient velocity to escape. In contrast, the color of the hot gas mass flux remains consistent on each side of the midplane. This is because the hot gas has been launched with high enough velocity to escape the simulation domain as a wind with nearly constant mass flux. However, the intermediate-temperature mass flux decreases as $|z|$ increases (color dilutes) with no clear evidence of inflow following each outflow, implying a transition to the warm phase due to a short cooling time. 

In the top panel of Figure~\ref{fig:mflux}, we present the time evolution of the SFR surface density measured from the mass of star clusters younger than 10~Myr. Star formation is bursty, involving an order of magnitude level fluctuations. However, the system approaches a quasi-equilibrium, self-regulated state, meaning that the average properties do not show a strong secular evolution. We select the time range of $t=200-550\Myr$ for analysis, covering many feedback cycles and outflow events so as to investigate the outflow properties in the statistical steady state.
The mean SFR surface density decreases by $20\%$ within this time range as the system continuously loses the gas mass by star formation and outflows.

Within the $t=200-550\Myr$ time range, we can identify 7-8 star formation peaks. However, from the mass flux evolution, we can observe only four clear breakouts (strong mass outflows in all phases). Two star burst events at $t\sim250-300\Myr$ do not result in strong outflows. We can see some hints of outflows for these events, but they fail to break out. This is because the large amount of material lifted up after the star formation burst at $t\sim 200\Myr$ falls back during the subsequent relatively weak star formation events at $t\sim 250$ and $300\Myr$.  On these and other occasions, the  conjunction of a ``returning'' fountain inflow with a subsequent starburst crushes and suppresses what might otherwise be a successful 
outflow.

\subsection{Time Averaged Properties}\label{subsec:tigress_tavg}
Despite large temporal fluctuations, the  overall evolution reaches a quasi-steady state and we can investigate the time-averaged quantities to understand mean behavior. We use simulation outputs between 200 and 550 Myr to construct time averaged profiles as
\begin{equation}\label{eqn:tavg}
\abrackets{q}_{\rm ph}(z) = \frac{\sum_{t} \overline{q}_{\rm ph}(z;t)\Delta t}{t_{\rm bin}},
\end{equation}
where $t_{\rm bin} = 350\Myr$ and $\Delta t = 1\Myr$ is the time interval of the output dump.

To understand kinematics and dynamics of outflowing gas, it is crucial to understand the contribution of each component in the momentum equation.\footnote{Note that, for simplicity, here we have dropped the Coriolis force and tidal potential terms arising from the galactic differential rotation included in the simulation. \citet{Kim&Ostriker18} presents the full equations, analyzes each term, and concludes that these terms have negligible impact on the outflows we are analyzing in this paper. Note also that on the RHS of Equation (14) in \citet{Kim&Ostriker18}, $\Phi_{\rm tot}\divergence{\rho\vel}$ is missing.} By taking horizontal and temporal averages, explicitly separating and summing over phases ``ph,'' we obtain the steady-state vertical momentum equation as
\begin{equation}\label{eqn:mom}
\begin{split}
 \deriv{}{z}\sum_{\rm ph} \abrackets{P_{\rm turb,z} + P_{\rm th} + \Pi_{\rm mag,z}}_{\rm ph} = \\
-\sum_{\rm ph} \abrackets{\rho\rbrackets{\deriv{\Phi_{\rm ext}}{z} +\deriv{\Phi_{\rm sg}}{z}}}_{\rm ph},   
\end{split}
\end{equation}
where the three  terms on the left hand side are the
turbulent, thermal, and magnetic force components,
respectively, and the
two terms on the right hand side are external and self gravity, respectively. $\Phi_{\rm ext}$ is the gravitational potential due to old stellar disk and dark matter halo as prescribed in Equation (\ref{eqn:pot}), and $\Phi_{\rm sg}$ is the gravitational potential of the gas  obtained by solving Poisson's equation ($\Phi_{\rm sg}$ includes the negligible contribution from young star clusters).  The thermal, $P_{\rm th }$, and turbulent, $P_{\rm turb,z}=\rho v_z^2$, stresses correspond to the thermal and turbulent vertical pressure,\footnote{We note that we use the term ``turbulent'' even though the motions of gas at high altitude are more or less ordered, dominated by either outflow or inflow at a given time that could be considered  ``ram pressure.'' However, inflows and outflows do coexist, and the dominance of one or the other alternates in time. Therefore, in a horizontally and temporally averaged sense the pressure arises from a variance of vertical velocity, and we simply adopt the nomenclature ``turbulent pressure'' for any Reynolds stress terms.  
} respectively. However, the magnetic stress,  $\Pi_{\rm mag,z}= (B_x^2+B_y^2-B_z^2)/(8\pi)$, is smaller than the magnetic pressure, $P_{\rm mag} = (B_x^2+B_y^2+B_z^2)/(8\pi)$, due to the magnetic tension term, $-B_z^2/(4\pi)$.

\begin{figure*}
	\centering
 	\includegraphics[width=\textwidth]{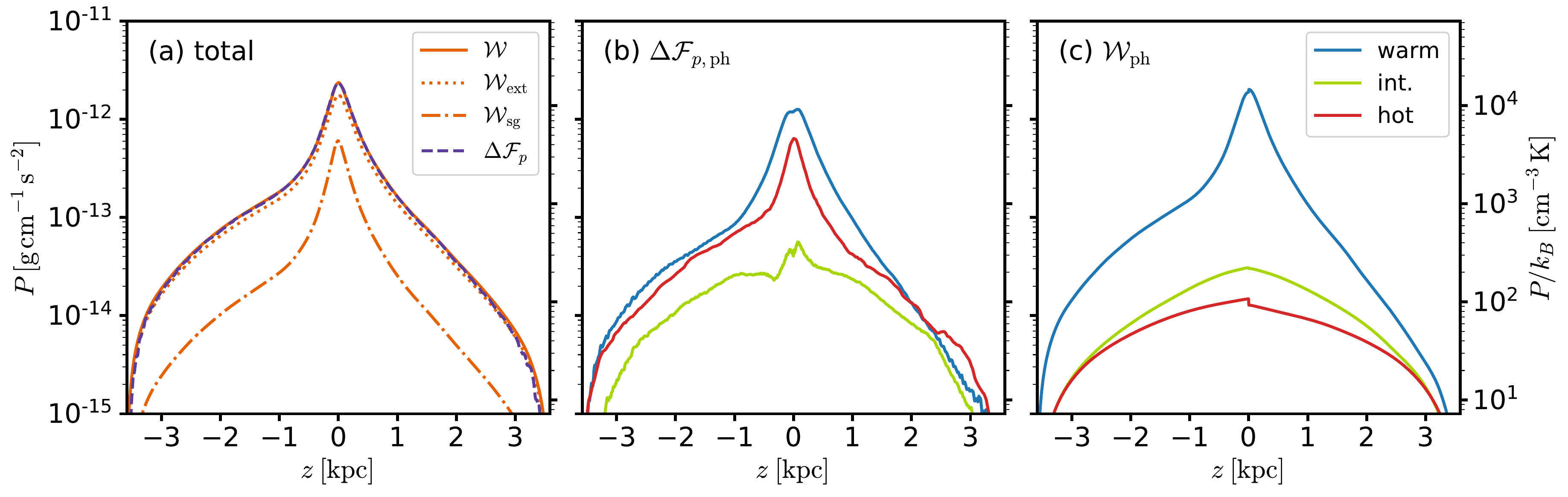}
	\caption{{\bf (a)} The total momentum flux difference (or vertical support; the LHS of Equation (\ref{eqn:mom2})) and the weight of gas (the RHS of Equation (\ref{eqn:mom2})) as a function of $z$. The almost perfect agreement between $\cal W$ and $\Delta {\cal F}_p$ in (a) demonstrates the validity of vertical dynamical equilibrium. The weight is further decomposed into the contribution from external (dotted) and self (dot-dashed) gravity terms, showing the dominance of the external gravity term. We also show  phase-separated contributions to {\bf (b)} the momentum flux differences and {\bf (c)} the weight.}
	\label{fig:vertical_equilibrium}
\end{figure*}

By integrating Equation (\ref{eqn:mom}) from the top/bottom of the simulation domain to the height $z$, we can rewrite the equation in terms of the momentum flux difference (or vertical ``support'') and the weight of gas
\begin{equation}\label{eqn:mom2}
\sum_{\rm ph} \sbrackets{ \Fph{p}{ph}(z)- \Fph{p}{ph}(\pm L_z/2)}
= \sum_{\rm ph} \mathcal{W}_{\rm ph}(z).
\end{equation}
Here the total flux of momentum (subscript ``p'') in a given phase at $z$ is defined by
\begin{equation}\label{eqn:Fp}
{\Fph{p}{ph}}(z) \equiv \abrackets{P_{\rm turb,z} + P_{\rm th} + \Pi_{\rm mag,z}}_{\rm ph}(z),
\end{equation}
and the total weight of gas in a given phase above $z$ is defined by
\begin{equation}\label{eqn:weight}
\mathcal{W}_{\rm ph}(z) \equiv \int_{\pm L_z/2}^z \abrackets{\rho\rbrackets{\deriv{\Phi_{\rm ext}}{z} +\deriv{\Phi_{\rm sg}}{z}}}_{\rm ph} dz \,.
\end{equation}

Figure~\ref{fig:vertical_equilibrium}(a) compares the momentum flux difference ($\Delta \Fmom$, dashed; the LHS of Equation~(\ref{eqn:mom2})) and the weight of the total gas ($\mathcal W$, solid; the RHS of Equation~(\ref{eqn:mom2})). On average, the vertical support equals the weight of the gas, meaning that the vertical dynamical equilibrium holds very well, as shown in previous work \citep[e.g.,][]{KOK13,KO15b}. This again justifies the quasi-steady state assumption. We also show that the gas weight is mostly due to the external gravity, especially above the gas scale height. 

In panels (b) and (c), we decompose the vertical support and the weight into different thermal phases. Since the warm gas dominates in terms of mass in the simulation, the weight is dominated by the warm medium everywhere, while the vertical support is provided by both warm and hot phases comparably, especially at high-$|z|$. By comparing the vertical support and weight for each phase, we note that the warm gas is lacking in support, while the hot gas shows a large excess in support. The support and weight of the intermediate phase are roughly balanced within the phase without significant excess or deficit.

\begin{figure*}
	\centering
    \includegraphics[width=\textwidth]{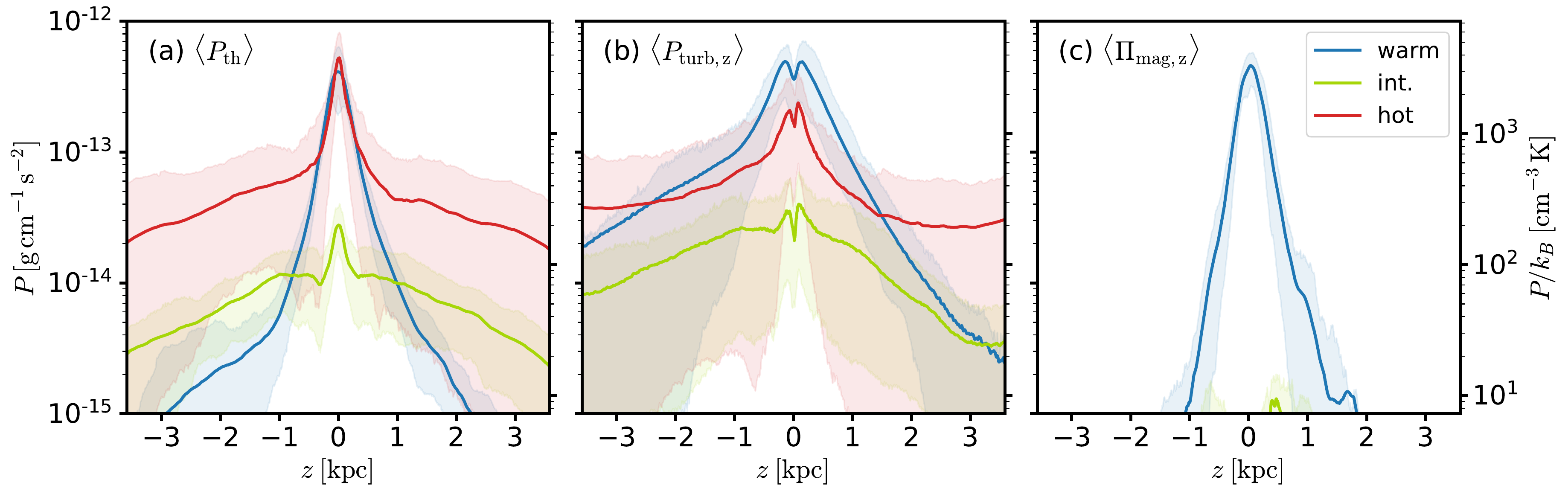}
	\caption{Horizontally and temporally averaged vertical profiles of {\bf (a)} thermal pressure $\abrackets{P_{\rm th}}$, {\bf (b)} turbulent pressure $\abrackets{P_{\rm turb,z}}$, and {\bf (c)} magnetic stress 
	$\abrackets{\Pi_{\rm mag,z}}$. The shaded area covers the 16th and 84th percentiles of temporal fluctuations.} 
	\label{fig:mean_profiles}
\end{figure*}

In order to investigate the momentum flux further, we construct averaged profiles of individual components and phases of the vertical stress. Figure~\ref{fig:mean_profiles} shows the horizontally and temporally averaged profiles of 
(a) thermal, (b) turbulent, and (c) magnetic stress terms for different thermal phases. Near the midplane, $|z|<H$, Figure~\ref{fig:vertical_equilibrium}(b) shows that the majority of support arises from the warm phase, and Figure~\ref{fig:mean_profiles} shows that the thermal, turbulent, and magnetic components of the warm phase stress are comparable to each other. The thermal pressure of the hot gas is also significant in this region (Figure~\ref{fig:mean_profiles}(a)). At high altitude, $|z|>(1-2)H$, the hot gas dominates both thermal and turbulent pressures. The turbulent pressure of the warm phase is also substantial, but drops rapidly as the warm outflow falls back at $|z|\sim1-2\kpc$ (see Figure~\ref{fig:mflux}). However, as shown in Figure~\ref{fig:vertical_equilibrium}(b), the ``support'' from hot and warm phases is similar in this region, since the support arises from the momentum flux ``difference''. This explains why, although the intermediate phase pressures are always negligible in terms of the absolute values, the vertical support is not negligible (note that the support of the intermediate phase is mostly compensated by its own weight, see Figure~\ref{fig:vertical_equilibrium}(b) and (c)). The magnetic component is negligible in the high-$|z|$ region so that we may safely ignore the magnetic field contribution in the following analysis.

\begin{figure*}
	\centering
    \includegraphics[width=\textwidth]{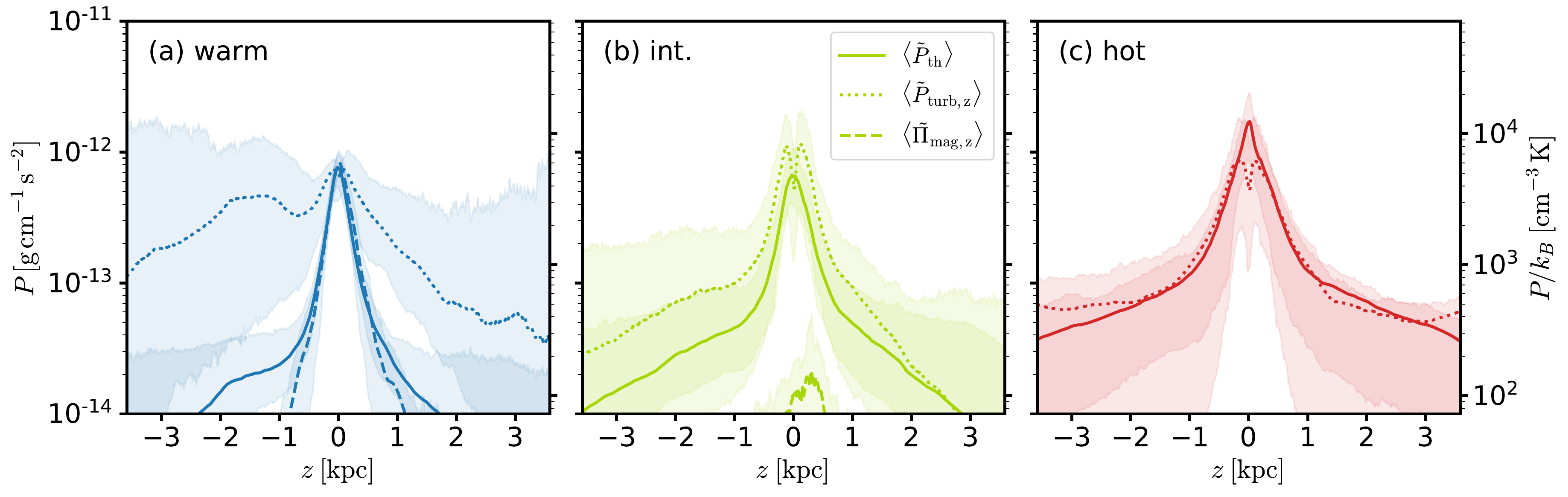}
	\caption{Vertical profiles of ``typical'' thermal pressure ($\tilde{P}_{\rm th}$; solid), turbulent pressure ($\tilde{P}_{\rm turb,z}$; dotted) and magnetic stress ($\tilde{\Pi}_{\rm mag,z}$; dashed) for the {\bf (a)} warm, {\bf (b)} intermediate, and {\bf (c)} hot  phases. The shaded area covers the 16th and 84th percentiles of temporal fluctuations. 
	Typical values in each phase are obtained by dividing horizontal averages by the area fraction of the phase, e.g. $\tilde{P}_{\rm th,ph}\equiv \abrackets{P}_{\rm th,ph}/f_{\rm A,ph}$.
	} 
	\label{fig:typical_profiles}
\end{figure*}

Figure~\ref{fig:typical_profiles} shows the ``typical'' vertical stress of each component for each phase, obtained by dividing the mean stress by the fraction of area occupied by each gas phase, $f_{\rm A,ph}\equiv\abrackets{1}_{\rm ph}$. For example, $\tilde{P}_{\rm th,ph}\equiv\abrackets{P_{\rm th, ph}}/f_{\rm A,ph}$. In contrast to Figure~\ref{fig:mean_profiles}, these profiles show the relative importance of each of the three stress components to the vertical dynamics for a particular phase. Interestingly, the typical turbulent pressure of the warm medium is the largest and dominates the other two at most heights. This figure reinforces the conclusions from Figure~\ref{fig:mean_profiles}: (1) for the warm phase, all thermal, turbulent, and magnetic components are comparable near the midplane, while turbulent pressure dominates at high altitude, (2) the hot gas has comparable thermal and turbulent pressures at all heights, and (3) magnetic support is negligible at high altitude.

\section{Ballistic Model of the Warm Outflow}\label{sec:ballistic}

\subsection{Ballistic Model}\label{subsec:bal_model}
To describe the evolution of the warm outflows in the extra-planar region (e.g., Figure~\ref{fig:slicenT}), we first consider the simplest model, namely ballistic motion consistent with the conservation of the mechanical energy: $v_z^2/2 + \Phi = \textrm{constant}$. The ballistic model assumes that each warm gas entity evolves independently and the change of its velocity is solely due to the gravity (no hydrodynamic interactions). Note that the external gravity dominates at the high-altitudes so that $\Phi\approx\Phi_{\rm ext}$ is a good approximation. Since $\Phi_{\rm ext}$ is known and fixed in time (see Equation (\ref{eqn:pot})), we can easily calculate the vertical velocity of the warm outflow at an arbitrary height, $z$, from the conditions at launching, $z=z_i$, as
\begin{equation}\label{eqn:bal_vf}
v_z (z) = \pm \sqrt{v_i^2 -2\rbrackets{\Phi (z) - \Phi (z_i)}}\,
\end{equation}
where $v_i = v_z(z_i)$ is the vertical velocity at launching. 

Since the outflowing gas in the simulation is not launched with a single velocity, it is more informative to consider a velocity PDF (v-PDF). In order to predict the mass-weighted v-PDF, $dM/dv$, at a height $z$, we consider the conservation of the mass flux for fluid elements at a given velocity. The mass of gas in a velocity range between $v$ and $v+\delta v$ is given by
\begin{eqnarray}
\delta M(v) = \deriv{M}{v}
\delta v,
\end{eqnarray}
while the total mass flux, $\rho v$, can be written as
$\delta M(v)v/(A\delta z)$
where $A\delta z$ is a volume of a thin slab that the gas passes through. Assuming mass flux conservation of material in a given velocity range as it travels from $z_i$ to $z_f$ and changes its velocity from $v_i$ to $v_f=v_z(z_f)$, we have 
\begin{eqnarray}\label{eqn:Mv}
\deriv{M}{v}\Big|_{v_f} v_f \delta v_f = \deriv{M}{v}\Big|_{v_i} v_i\delta v_i\,.
\end{eqnarray}
From Equation (\ref{eqn:bal_vf}), we have $v_f \delta v_f = v_i\delta v_i$, so that  Equation (\ref{eqn:Mv}) simply becomes
\begin{eqnarray}\label{eqn:vpdf}
\deriv{M}{v}\Big|_{v_f} = \deriv{M}{v}\Big|_{v_i}\,.
\end{eqnarray}
Therefore, under the assumption of the mass flux conservation, 
the v-PDF at $z_{f}$ can be obtained via a velocity shift applied to the initial 
v-PDF at $z_{i}$ according to the ballistic equation (Equation (\ref{eqn:bal_vf})).

\subsection{Comparison Between Simulation and Model}\label{subsec:bal_comparison}
\begin{figure*}
	\centering
	\includegraphics[width=\textwidth]{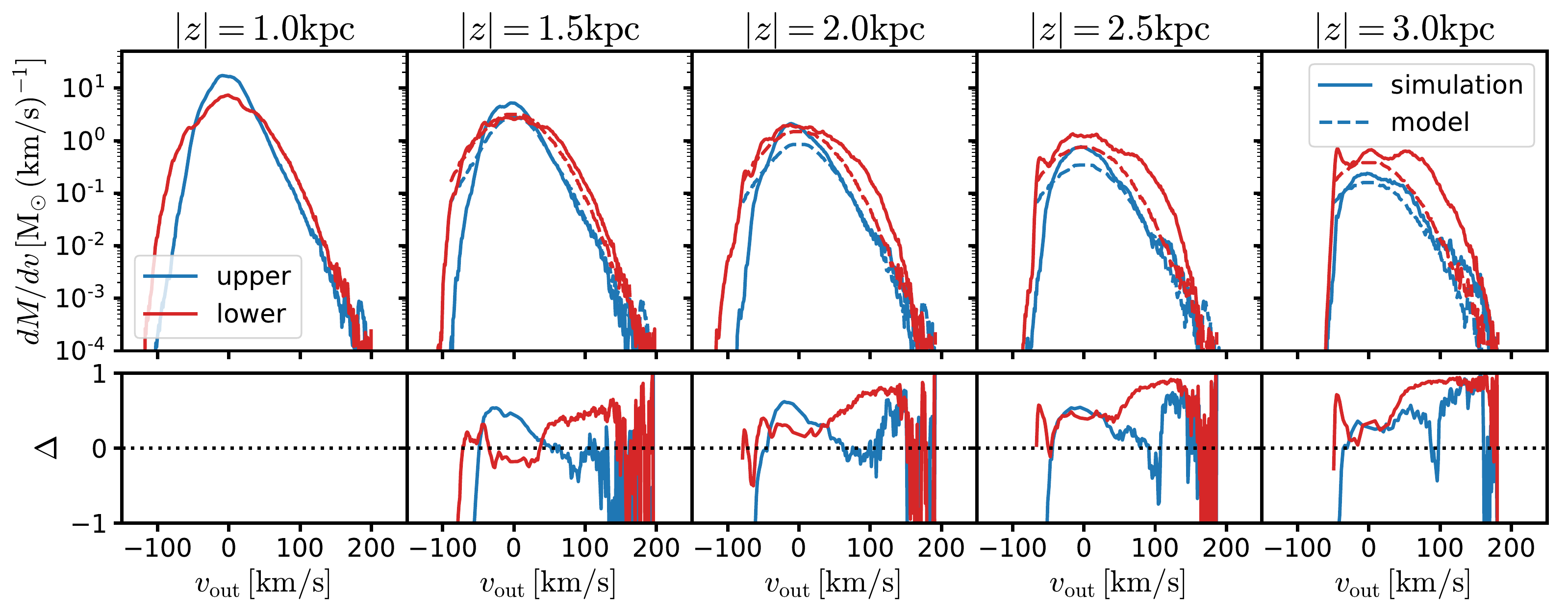}
	\caption{The mass-weighted velocity PDFs for the warm phase along with the fractional difference between the simulation data and the ballistic model, $\Delta=$ (Simulation - Model)/Simulation in the bottom. From left to right, we show the v-PDFs from the simulation data at $|z|=1$, 1.5, 2, 2.5, and 3~kpc as solid lines for the upper (blue) and lower (red) half of the disk. Note that the vertical velocity is multiplied by the sign of $z$ to convert it to the outflow velocity. For $|z|>1\kpc$, we plot the ballistic model prediction (see Section~\ref{subsec:bal_model}) as the dashed line adopting the v-PDF at $|z|=1\kpc$ as an initial condition.
 } 
	\label{fig:ballistic_warm}
\end{figure*}

\begin{figure*}
	\centering
	\includegraphics[width=\textwidth]{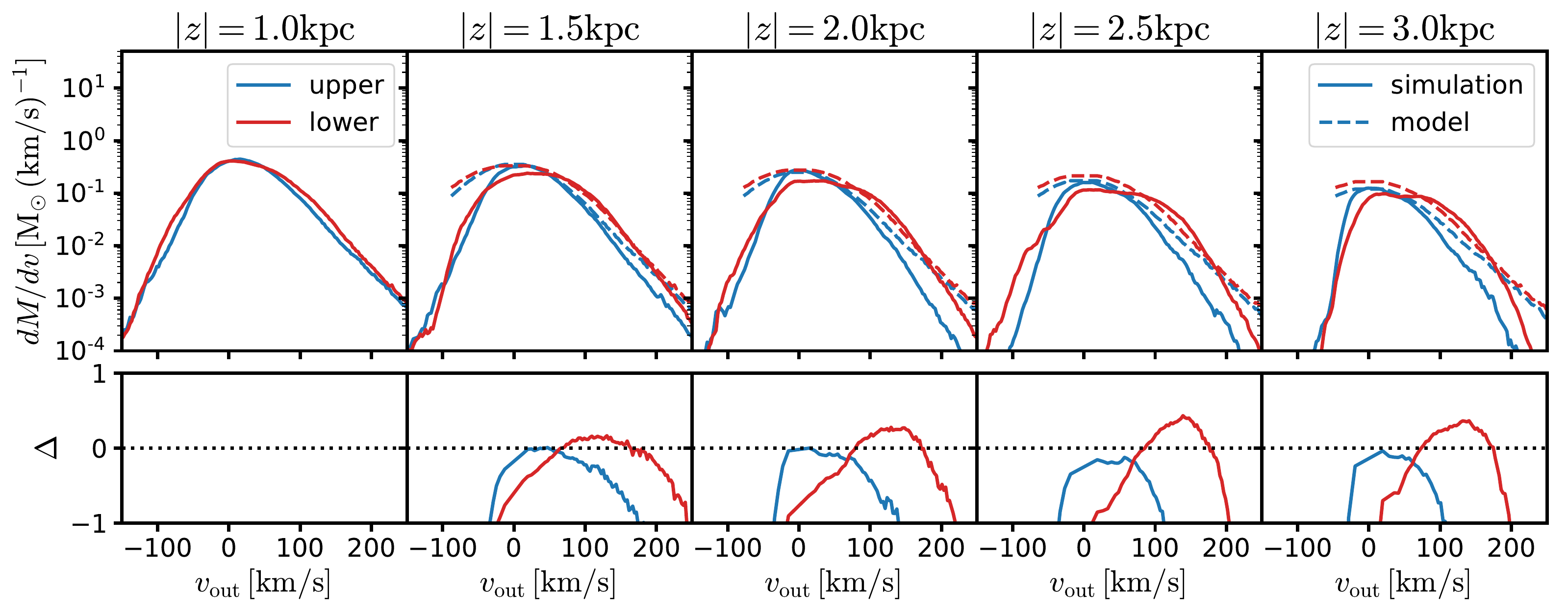}
	\caption{Same as Figure~\ref{fig:ballistic_warm}, but for the intermediate phase.} 
	\label{fig:ballistic_int}
\end{figure*}

\begin{figure*}
	\centering
	\includegraphics[width=\textwidth]{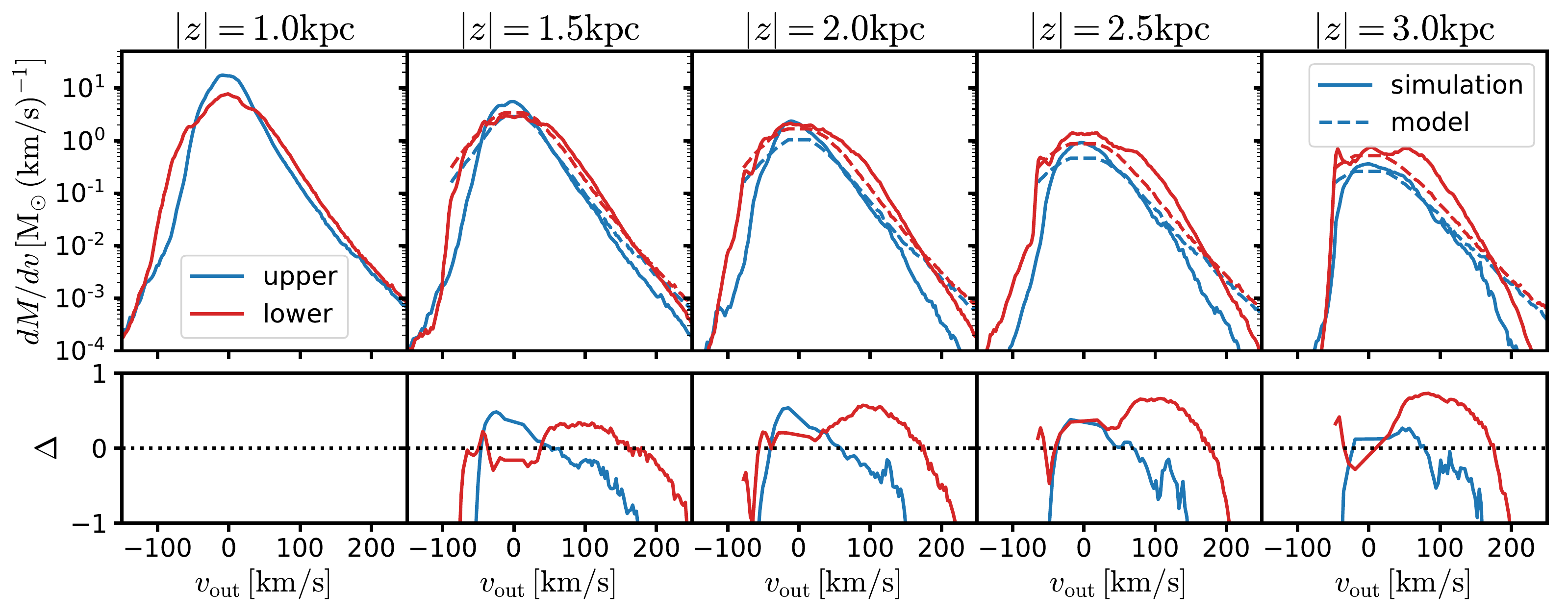}
	\caption{Same as Figure~\ref{fig:ballistic_warm}, but for the sum of the warm and intermediate phases.} 
	\label{fig:ballistic_warm_int}
\end{figure*}

Since the typical turbulent pressure of the warm phase is the largest among all the momentum flux terms and phases (see Figure~\ref{fig:typical_profiles}), the ``zeroth order'' expectation is for the warm outflow to evolve more or less ballistically, unaffected by 
self-interactions\footnote{With self-interactions, collisions of fluid elements with different velocities would yield a fluid element at intermediate velocity. To the extent that the volume filling factor is sufficiently low (see Section \ref{sec:flux_Rcl}), self interactions can be neglected.} or interactions with other phases. Therefore, the warm phase is the most suitable component to be compared with the ballistic model.

We compare the mass-weighted v-PDFs of the warm phase as a function of height obtained from the simulation with those predicted by the ballistic model (Section~\ref{subsec:bal_model}). For simulation snapshots, we first calculate the mass-weighted v-PDF at each height $z$ and take time averages. For the ballistic model, we use the v-PDF at $|z_i|=1\kpc$ from the simulation as an initial (launching) condition and calculate the predicted v-PDF at heights $|z_f|=1.5$, 2, 2.5, and 3$\kpc$ based on Equations (\ref{eqn:bal_vf}) and (\ref{eqn:vpdf}). Note that we treat the upper ($z>0$) and lower ($z<0$) sides of the simulation domain independently. Assuming steady injection of the outflowing gas through the $z=z_i$ plane, both outflowing and inflowing components of the v-PDF at different heights can be predicted (see $\pm$ signs in Equation (\ref{eqn:bal_vf})). Due to the limited vertical extent of the simulation domain, however, outflowing gas with sufficiently high speed is likely to exit the simulation box. Thus, for a fair comparison with the simulation, we set a cut-off velocity in the computation of the inflowing component, 
\begin{eqnarray}\nonumber
|v_{\rm{cut}}| \equiv \sqrt{2\, \sbrackets{ \Phi(z_i) - \Phi(\pm L_z/2)}}=98\kms\,,
\end{eqnarray}
where $L_z$ is the vertical size of the simulation box.
 
Figure~\ref{fig:ballistic_warm} shows the comparison between simulation and model for the warm phase at heights above (blue) and below (red) the galactic plane, respectively.
The quantitative comparison between the simulation data (solid) and the ballistic predictions (dashed) is presented as the fractional difference in the respective bottom panels. The fractional difference is defined by 
\begin{equation}\label{eqn:delta}
\Delta\equiv \frac{dM/dv({\rm simulation}) - dM/dv({\rm model})}{dM/dv({\rm simulation})}.
\end{equation}
The positive or negative $\Delta$ means that the ballistic model under- or over-predicts the mass in a velocity bin, respectively.

Despite the highly simplistic assumptions applied here, the ballistic model generally recovers the v-PDF at different heights quite reasonably. In order to construct the v-PDF from the ballistic model, we also neglect interaction between outflow and inflow for a given phase, which certainly exists due to the bursty nature of the star formation and outflows (see Figure~\ref{fig:mflux}). Therefore, the overall agreement of the ballistic prediction to the simulation data gets worse as the distance between heights of launching and prediction gets larger and at higher outflow velocity bins. 

The ballistic model underestimates the mass of the warm phase at high velocity bins (positive $\Delta$ in Figure~\ref{fig:ballistic_warm}), without overestimation at low velocity bins (no significant negative $\Delta$ in Figure~\ref{fig:ballistic_warm}). This potentially means that the high velocity excess of the warm medium is not due to the acceleration of low velocity warm outflow itself but due to the addition of high velocity gas from the other phases. Considering the short cooling time of the intermediate phase, it is natural to expect a transition from the intermediate phase to the warm phase. For the intermediate phase, the typical cooling time above $|z|>1\kpc$ is $t_{\rm cool}\equiv \abrackets{P_{\rm th}}/[(\gamma-1)\abrackets{\mathcal{L}}]\sim$ few $\Myr$ (at $z=1-2$ kpc),
where $\mathcal{L}$ is the net volumetric cooling rate, increasing from low-altitude to high-altitude. This is shorter than or comparable to the outflow crossing time for the simulation domain from $|z|=1\kpc$ to $|z|=L_z/2$, $t_{\rm cross}=(L_z/2 - 1\kpc)/v_{\rm out} = 25\Myr \,(v_{\rm out}/100\kms)^{-1}$.

To further test the idea that cooling of intermediate-temperature gas produces high-velocity warm gas at large $z$, we perform the same comparison for the intermediate phase in Figure~\ref{fig:ballistic_int}. 
Here, we do not anticipate the general validity of the ballistic model for the intermediate phase since the non-negligible cooling in the intermediate phase violates the necessary assumptions for the ballistic model (mass conservation). Note that, although the phase transition due to cooling of the intermediate phase also means addition of mass to the warm phase, the mass contribution from the intermediate phase is not dominant for the warm phase so that the ballistic assumption we have made for the warm phase may still hold approximately (see Section~\ref{subsec:flux_simulation} for details). Bearing these caveats in mind, Figure~\ref{fig:ballistic_int} shows general deficits of the mass at high-velocity bins of intermediate-temperature gas (negative $\Delta$), without significant excesses in any other velocity bins. 
Since the mass fraction at high-velocity bins is smaller, the excess would be more prominent at these bins.

If we simply assume that the gas at the intermediate temperature cools and turns into the warm phase, the phase transition just moves the mass at a given velocity bin from one phase to the other. As shown in Figure~\ref{fig:ballistic_warm_int}, when we consider warm and intermediate phases together, the agreement is only partly improved. In particular, the excess of the warm phase on the upper side of disk at $50\kms<v_{\rm out}<150\kms$ is well counterbalanced by the deficit of the intermediate phase. However, on the lower side of disk, the intermediate phase does not make the ballistic model better.

So far, we have neglected any dynamical interaction between phases. However, the v-PDFs shown in Figures~\ref{fig:ballistic_warm} -- \ref{fig:ballistic_warm_int}  possess a signature of phase interaction. The outflow is generally asymmetric (see Figure~\ref{fig:mflux}), and this is evident in Figure~\ref{fig:ballistic_warm}, where the v-PDFs of the warm phase at $z=\pm 1\kpc$ show significant difference. In contrast, the intermediate phase shows very similar v-PDFs at $|z|=1\kpc$ for both sides. The asymmetry in the intermediate phase emerges as it moves outward and interacts with different warm phase outflows. In addition to the failure of the simple cooling idea, this clearly implies that the dynamical interaction between phases exists and affects the evolution of outflows noticeably. We quantify this effect in the next section.

\section{Mass, Momentum, and Energy Transfers Between Phases}\label{sec:flux_analysis}
\begin{figure*}
	\centering
	\includegraphics[width=0.9\textwidth]{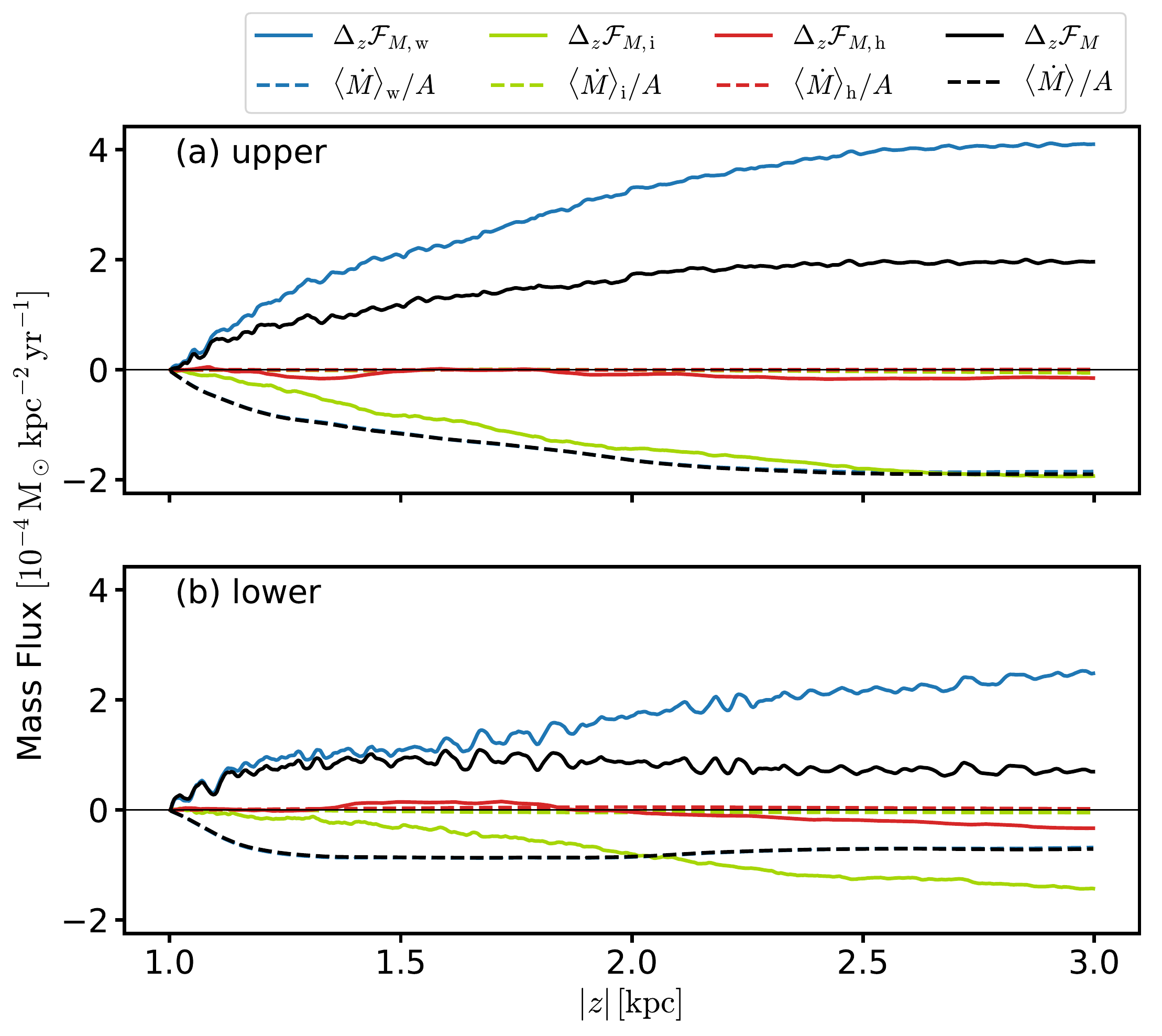}
	\caption{Each term of Equation (\ref{eqn:mcons}), showing the rate of mass change per area ($\dot{\abrackets{M}}/A$; dashed) and the flux difference ($\Delta_z \mathcal{F}_{\rm M}$; solid) for each phase (blue for warm, lime green for intermediate, and red for hot) as well as whole gas (black). Note that the mass change over the time period we consider is negligibly small for the intermediate and hot phases and  significant for the warm phase, $\dot{\abrackets{M}}\sim\dot{\abrackets{M}}_{\rm w}\gg\dot{\abrackets{M}}_{\rm i/h}$, making the blue dashed line invisible under the dashed black line.  }
	\label{fig:mfluxterms}
\end{figure*}

\begin{figure*}
	\centering
	\includegraphics[width=0.9\textwidth]{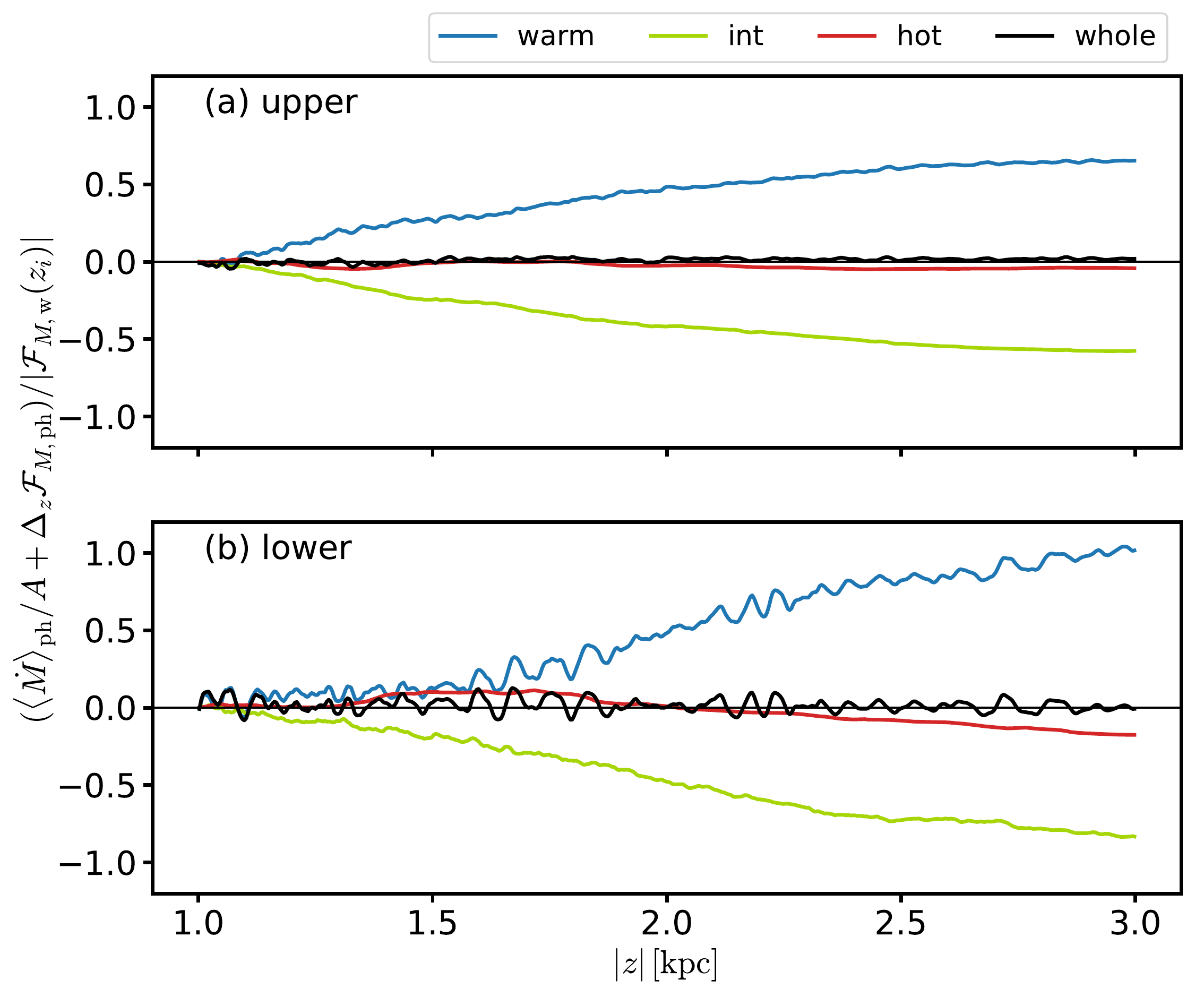}
	\caption{Net mass 
	gain per unit area per unit time 
	in each thermal gas phase ($\dot{\abrackets{M}}_{\rm ph}/A + \Delta_z \mathcal{F}_{\rm M,ph}$) between the height of interest $z$ and launching $z_i=1\kpc$. Separately  shown  are the {\bf (a)} upper and {\bf (b)} lower sides of the disk. All terms are normalized by the mass flux of the warm phase at $z=z_i$. The net mass 
	gained by the warm phase matches the mass flux lost by the intermediate-$T$ phase, 
	$\dot{\abrackets{M}}_{\rm w}/A + \Delta_z \mathcal{F}_{\rm M,w} \approx - \Delta_z \mathcal{F}_{\rm M,i}$. }
	\label{fig:mcons}
\end{figure*}
\begin{figure*}
	\centering
	\includegraphics[width=0.9\textwidth]{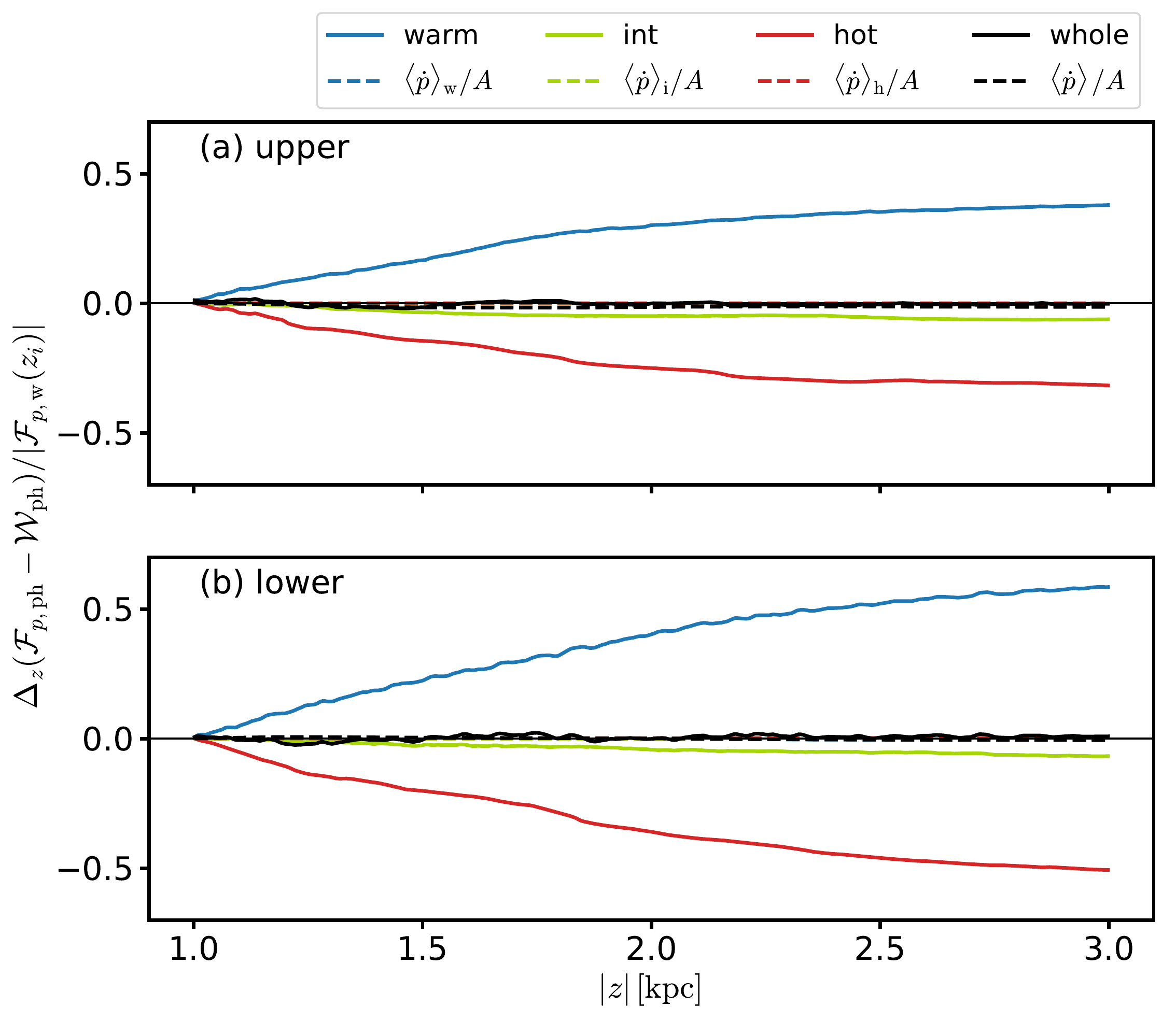}
	\caption{Same as Figure~\ref{fig:mcons}, but for the net momentum flux difference ($\Delta_z (\mathcal{F}_{\rm p,ph}-\mathcal{W}_{\rm ph})$; solid)  taking into account the flux loss due to climbing out of the potential; also shown is the momentum change rate per unit area ($\dot{\abrackets{p}}_{\rm ph}/A$; dashed). Note that the momentum changes are negligibly small for each phase individually as well as the whole gas. Although the mass change rate is non-negligible for the warm phase (see Figure~\ref{fig:mfluxterms}), its momentum change is small due to the preferentially low velocity of the warm phase (see Figure~\ref{fig:ballistic_warm}). }
	\label{fig:pcons}
\end{figure*}
\begin{figure*}
	\centering
	\includegraphics[width=0.9\textwidth]{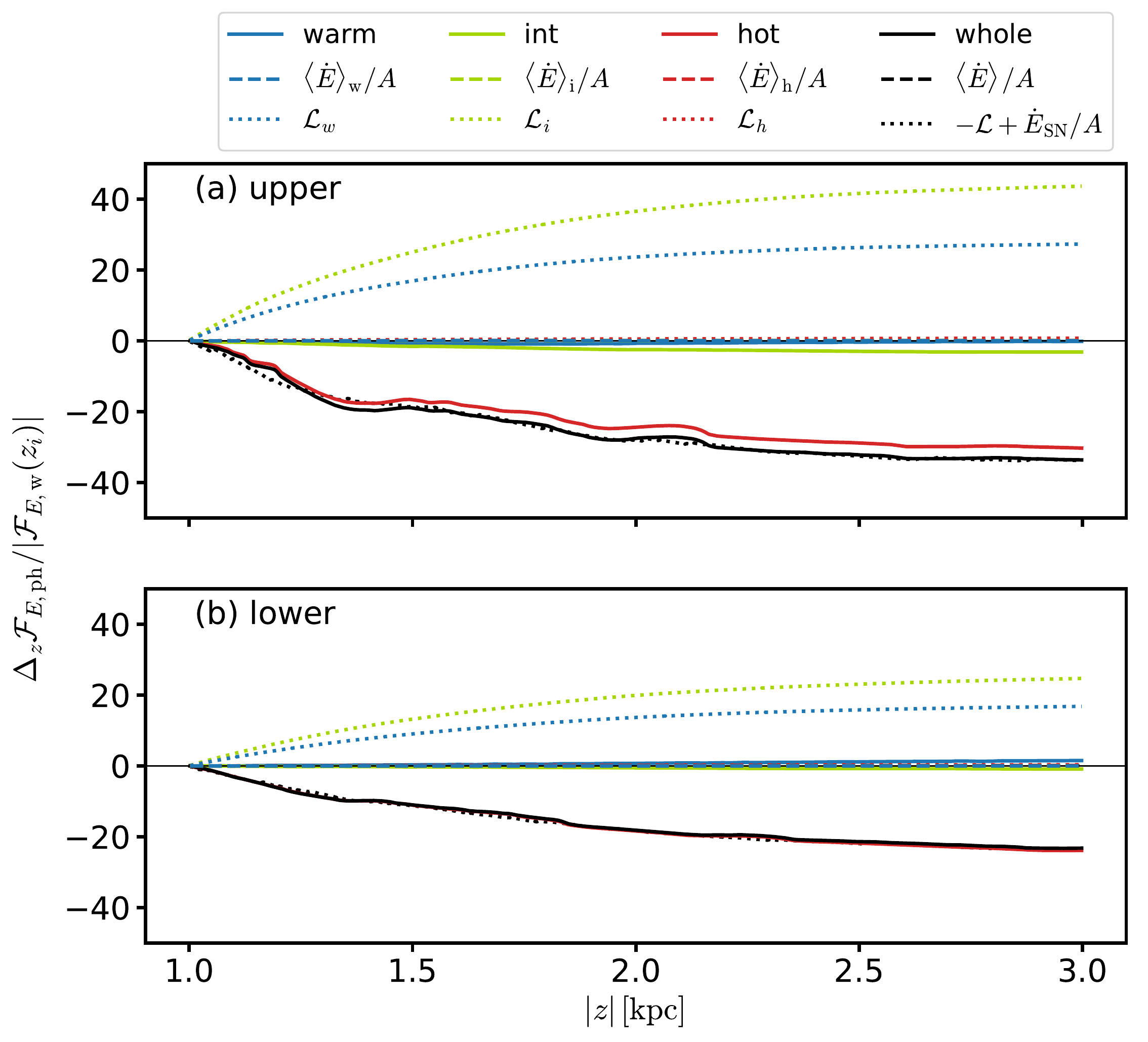}
	\caption{Same as Figure~\ref{fig:mcons}, but for the energy flux difference. In addition, the net cooling rate per unit area of each thermal gas phase is shown as dotted lines. The time dependent energy changes in  the volume ($\dot{\abrackets{E}}_{\rm ph}/A$; dashed) are negligible. The black dotted line denotes the total energy source term of the energy equation including the net cooling and direct SN energy injection; this net loss (negative source) has an excellent match to the net energy flux change of the whole medium, most of which is loss of hot gas energy flux. Note that the range of the y-axis is about two orders of magnitude larger than that in Figures~\ref{fig:mcons} and \ref{fig:pcons}, demonstrating that a large amount of energy is brought into the extra-planar region via hot-phase gas, but this is mostly lost to cooling.}
	\label{fig:econs}
\end{figure*}

In order to provide a more complete picture of multiphase outflows and interaction between phases, we now analyze key terms in the hydrodynamics equations of mass, momentum (in the $z$ direction), and energy conservation (we neglect magnetic fields in this section as we already saw that the magnetic terms are negligible in outflows; see Section~\ref{subsec:tigress_tavg}). We first take the horizontal average for each thermal phase and the temporal average for time range of $t\in(t_1,t_2)$ as defined in Equations (\ref{eqn:havg}) and (\ref{eqn:tavg}). Then, we further integrate the equations outward along the vertical direction from $z_i$ to $z$, where $z_i=\pm$ 1kpc. As we shall show in Section~\ref{subsec:flux_simulation}, we treat upper and lower halves of the disk separately. 

The set of hydrodynamic partial differential equations yields integrated relationships based on the three conservation laws for mass, momentum, and energy:

\begin{equation}
\sum_{\rm ph}\sbrackets{\dot{\abrackets{M}}_{\rm ph}/A + \Delta_z\Fph{M}{ph}}= 0,
\label{eqn:mcons}
\end{equation}
\begin{equation}\label{eqn:pcons}
\sum_{\rm ph} \sbrackets{\dot{\abrackets{p}}_{\rm ph}/A + 
 \rbrackets{ \Delta_z\Fph{p}{ph} - \Delta_z\mathcal{W}_{\rm ph}}} =0,
\end{equation}
\begin{equation}\label{eqn:econs}
\sum_{\rm ph} \sbrackets{\dot{\abrackets{E}}_{\rm ph}/A +
\Delta_z\Fph{E}{ph}}= -\mathcal{L}_{\rm ph}+ \dot{E}_{\rm SN}/A.
\end{equation}

The rate of change in mass within the volume of interest, $\dot{\abrackets{M}}_{\rm ph}$, is defined by
\begin{equation}\label{eqn:mdot}
    \dot{\abrackets{M}}_{\rm ph} \equiv \sum_{z'=z_i}^{z}
    \abrackets{\pderiv{\rho (z')}{t}}_{\rm ph}A\Delta z ,
\end{equation}
where $\Delta z$ is the cell size and $A=L_xL_y$ is the total area of the horizontal plane. Similarly, $\dot{\abrackets{p}}_{\rm ph}$ and $\dot{\abrackets{E}}_{\rm ph}$ are defined by replacing $\rho$ in  Equation (\ref{eqn:mdot}) with $\rho v_z$ and by $\rho v^2/2+P_{\rm th}/(\gamma-1)$, respectively, where $\gamma=5/3$, is the adiabatic index.

The momentum flux  $\Fph{p}{ph}$ in phase ``ph'' is defined by Equation~(\ref{eqn:Fp}), while mass and energy fluxes are respectively defined by
$\Fph{M}{ph}\equiv \abrackets{\rho v_{\rm out}}_{\rm ph}$ 
and $\Fph{E}{ph}\equiv \abrackets{\rho v_{\rm out}\mathcal{B}}_{\rm ph}$,
where the Bernoulli parameter is 
\begin{equation}\label{eqn:Bernoulli}
\mathcal{B} \equiv \frac{v^2}{2} + \frac{\gamma}{\gamma -1} \frac{P_{\rm th}}{\rho} + \Phi.
\end{equation}
The weight of gas $\mathcal{W}_{\rm ph}$ is defined by Equation (\ref{eqn:weight}). 
For the mass, momentum, and energy fluxes, and for the weights, $\Delta_z q\equiv   q(z)-q(z_i)$, i.e. the difference between the value at the height of interest $z$ and the  outflow launching point $z_i$.   

The source terms in Equation (\ref{eqn:econs}) indicate
the energy loss by cooling $\mathcal{L}_{\rm ph}\equiv \sum_{z'=z_i}^{z}\abrackets{n^2\Lambda(T)-n\Gamma}_{\rm ph}(z')\Delta z$ and the gain by 
direct SN energy injection $\dot{E}_{\rm SN} \equiv 10^{51}\erg\,N_{\rm SN}/t_{\rm bin}$. 

The first term on the LHS of Equations (\ref{eqn:mcons}), (\ref{eqn:pcons}), and (\ref{eqn:econs}) is the change of mass, momentum, and energy in each thermal phase within the volume and time interval of interest. If the system is in a perfect steady state, the summation over thermal phases would be zero. Due to the dynamic and bursty nature of the simulation, the steady state assumption within the volume we are analyzing is not always satisfied even after the long temporal averaging ($t_1=200\Myr$ and $t_2=550\Myr$). We thus keep this term to demonstrate how significant the unsteady behavior is. The total mass difference can be particularly large compared to the mass flux term. Although we consider the volume far from the midplane $|z|>1\kpc$, there is still direct SN energy injection due to SNe from runaways, which serves as a source term in the energy equation.

\subsection{Simulation Results\label{subsec:flux_simulation}}

In order to understand mass exchange between phases, we plot individual terms of Equation (\ref{eqn:mcons}) in Figure \ref{fig:mfluxterms}.\footnote{We would like to note that our finding that the warm mass flux increase with $z$ might seem in apparent contradiction with what is shown in Figure 7 of \citet{Kim&Ostriker18}. There, the mass fluxes of the warm gas during the outflow-dominated (Fig.~7a) and inflow-dominated (Fig.~7b) periods are analyzed separately, yielding the result that the absolute value of the mass flux decreases with $z$ in both periods. Here, we analyse the mass flux over the entire selected temporal range without making any distinction between outflow-dominated and inflow-dominated periods. Thus, the mass flux here ($\Fph{M}{w}\equiv \abrackets{\rho v_z}_{\rm w}$) should be seen as the difference between the positive mass flux during the outflow-dominated period and the negative value during the inflow-dominated period in \citet{Kim&Ostriker18}.}
In absence of interactions between the phases, we would expect that increase (decrease) in mass flux for each phase would be balanced by an equivalent decrease (increase) in $\dot{\abrackets{M}}$ for each phase. In other words, each phase would individually satisfy Equation (\ref{eqn:mcons}) (solid and dashed lines in Figure~\ref{fig:mfluxterms} would  compensate each other color by color). This is not the case. 

For the hot phase (red), the mass flux difference and the corresponding $\dot{\abrackets{M}}_{\rm h}$ are both nearly zero, showing that the hot gas coming in from $z_i$ simply goes out without having significant mass exchange interactions with either of the two phases. The intermediate phase (lime green) has $\dot{\abrackets{M}}_{\rm i}$ negligible, while its mass flux difference (divergence) is substantial and negative. This means that the intermediate phase gas that comes in from the bottom of the volume neither escapes outward nor stays in the volume to increase the mass of the intermediate-$T$ gas. This points to the fact that mass exchange must exist between the warm and intermediate phases. 

Within the time interval considered, the mass flux difference (divergence) of the warm phase is positive.  This could correspond either to net flow outward through the top of the box (into the CGM), or inward  through the bottom of the box (a falling fountain)\footnote{Since warm gas velocities are usually insufficient to escape from the galaxy, a positive divergence in the flux would generally be expected to represent an inward fountain flow. However, the vertical extent of this simulation domain is limited, so that a strong burst can sometimes lead to substantial net outflow for the warm phase. This in fact occurs at $\sim 250\Myr$ in the lower half (see Figure~\ref{fig:mflux}).} The total mass change within the period  is solely due to the warm phase (blue dashed is almost perfectly overlaid with black dashed) and smaller than the mass flux difference (magnitude of blue solid is larger than blue/black dashed). More warm gas flows out of the volume than the rate of change of the stored warm mass, necessitating a transfer of mass flux from the intermediate to the warm gas.

In short, Figure~\ref{fig:mfluxterms} shows three robust features of the mass flux difference: (1) the hot gas mass flux is nearly constant with height, $\Delta_z \Fph{M}{h} \sim 0$, (2) the intermediate gas mass flux decreases with height, $\Delta_z \Fph{M}{i} < 0$, without producing an increase in the mass $\dot{\abrackets{M}}_{\rm i}\sim 0$, and (3) the warm gas mass flux increases with height, $\Delta_z \Fph{M}{w} > 0$.  The increase in warm mass flux with height can be attributed to a combination of flux transfer from the intermediate phase and a reduction in the ``stored'' warm mass over the period.  

To directly compare the mass flux loss and gain within each phase, in Figure \ref{fig:mcons} we show the net mass flux gain per unit area per unit time of each phase $\dot{\abrackets{M}}_{\rm ph}/A + \Delta_z \Fph{M}{ph}$, normalized by the mass flux at $|z_i|=1\kpc$ of the warm phase  $|\Fph{M}{w} (z_i)|$.

Over the height range we consider here, the warm phase gains mass flux from the intermediate phase nearly continuously.  
Between 1 and 3 kpc, the decrease in the intermediate-$T$ flux, $\Delta_z \Fph{M}{i}$, and corresponding increase in the warm mass flux, amounts to $50$-$100\%$ of the launch value of $\Fph{M}{w}$ (at $|z|=1\kpc$).

Next, we investigate momentum exchanges between phases. Figure~\ref{fig:pcons} plots the ``net'' momentum flux difference $\Delta_z(\Fph{p}{ph} - \mathcal{W}_{\rm ph})$ rather than the momentum flux itself, which always decreases as the outflow climbs up the gravitational potential well, i.e.  $\Delta_z\Fph{p}{ph} < 0$ (see Figure \ref{fig:vertical_equilibrium}). We also plot $\dot{\abrackets{p}}_{\rm ph}/A$ to show that these terms are negligible and hence do not contribute significantly to momentum balance among the phases. The total net momentum flux difference is nearly zero since vertical equilibrium holds at every height, as we see in Figure~\ref{fig:vertical_equilibrium}. However, a significant momentum flux loss occurs in the hot phase, which clearly corresponds to a gain in the warm phase. Although there is net loss in the momentum flux of the intermediate phase as well (mainly due to direct phase transition as seen in Figure~\ref{fig:mcons}), the amount of the momentum flux transferred from the intermediate phase to the warm phase is negligible compared to that from the hot phase. The warm phase gains about $50\%$ of its original momentum flux, which is comparable to the loss from the hot phase. The momentum flux of the warm phase is dominated by the turbulent term, while the turbulent and thermal terms of the hot phase are similar (see Figure \ref{fig:typical_profiles}).

Lastly, Figure~\ref{fig:econs} plots the energy flux differences along with the cooling, SN energy injection rate, and rate of change of energy densities ($\dot{\abrackets{E}}_{\rm ph}/A$), all normalized to $|\Fph{E}{w} (z_i)|$. $\dot{\abrackets{E}}_{\rm ph}$ (dashed lines) is negligible for all the phases, so energy balance is maintained between flux differences (solid lines) and source terms (dotted lines; cooling in warm and intermediate and direct SN energy injection from runaways).  The hot gas energy flux drops with height ($\Delta_z \Fph{E}{h}<0$, solid red) but there is no significant increase of the energy flux in either the warm or intermediate phase.   Instead, cooling losses from these two phases (dotted blue and  green) are responsible for draining the hot gas energy, as well as for radiating away the additional SN energy injected within the volume.  That is, $\dot{E}_{\rm SN}/A- \Delta_z \Fph{E}{h} \approx (\mathcal{L}_{\rm w}+\mathcal{L}_{\rm {i}})$. 
The net loss of energy is large compared to the energy flux of the warm phase at $z_i$.

In summary, from Figures~\ref{fig:mcons}--\ref{fig:econs}, we conclude that the warm phase gains mass flux from the intermediate phase and momentum flux from the hot phase. 
The energy flux available in the hot gas is enormous, but there are significant losses  (20-30 times the original energy  flux in the warm medium) over $\Delta z=$2~kpc. Energy transfer between the hot and warm phases can occur by mixing or shocks, and substantial amount of energy imparted by these routes is lost from the simulation rather than  appearing in other phases due to very efficient radiative cooling in both warm and intermediate phases.

Using the net mass and momentum exchanges, we can understand the ballistic model results presented in Section~\ref{subsec:bal_comparison}. The phase transition from the intermediate phase to the warm phase indeed occurs and is substantial in terms of mass flux. However, the momentum gain from the intermediate phase is insignificant, implying that the phase transition from the intermediate phase occurs preferentially at low velocities (or includes both outflowing and inflowing components).
This is reason why the ballistic approximation of the warm phase works relatively well even though there is non-negligible mass transfer from the intermediate phase. The excess of the high velocity component in the warm phase seen in Figure~\ref{fig:ballistic_warm_int} comes from interaction with the hot phase, as demonstrated from momentum flux changes in Figure~\ref{fig:pcons}. 
Overall, interaction between phases results in a factor of 1.5-2 increase in the mass and momentum fluxes of the warm phase over $\Delta z=2\kpc$ compared to its steady-state injection fluxes.
We note that the hot medium sees 
a $40$-$50\%$ reduction in both kinetic ($\rho v_z^2$) and thermal ($P$) momentum fluxes  over the region studied.  

\subsection{Effective Size of Warm Clouds}\label{sec:flux_Rcl}
\begin{figure*}
	\centering
 	\includegraphics[width=\textwidth]{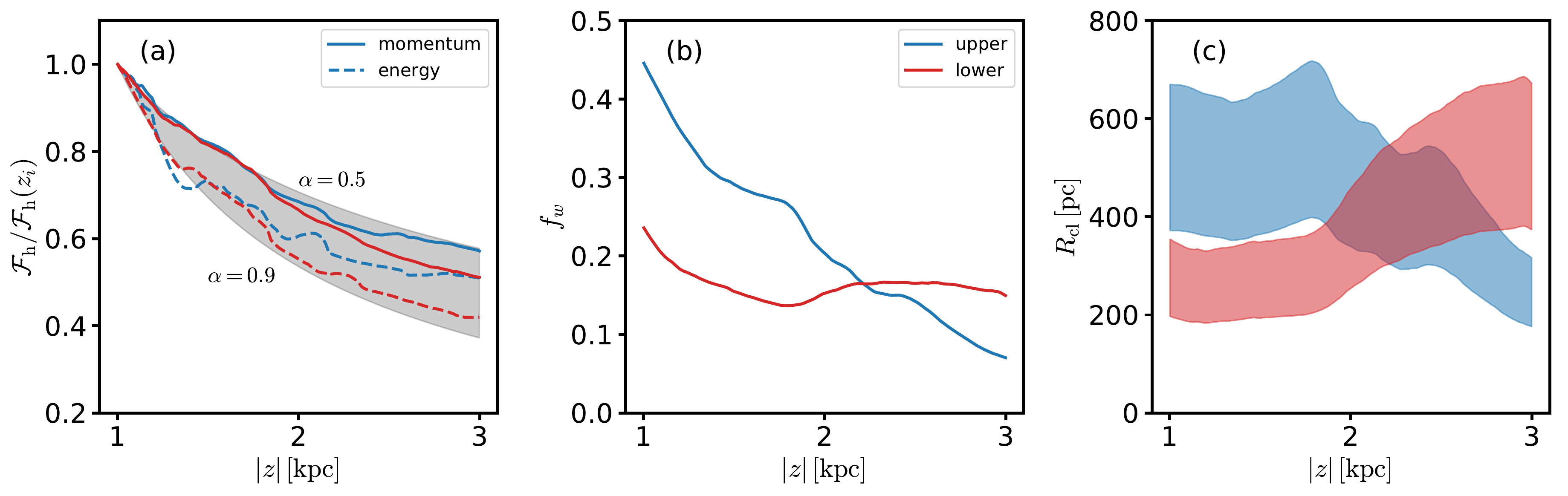}
	\caption{{\bf (a)} Normalized flux profiles of momentum (solid) and energy (dashed) for the upper (blue) and lower (red) sides of the disk. Grey shaded region encloses a simplified model $(z/z_i)^{-\alpha}$ with a range of $\alpha=0.5$ to 0.9. 
	{\bf (b)} Volume filling factor profiles of the warm phase.
	{\bf (c)} Effective size of warm clouds $R_{\rm cl}$ derived from Equation \ref{eqn:Rcl}. The shaded area covers the range of $\alpha$ we adopt in the panel (a), with larger $R_{\rm cl}$ corresponding to smaller $\alpha$. Note that the effective size of clouds is generally large, implying that in reality  the warm phase is likely to interact with the hot winds as a single entity rather than many small cloudlets.}
	\label{fig:rcl}
\end{figure*}

Figure~\ref{fig:pcons} shows that the net  momentum flux gain of the warm medium with increasing  $|z|$ (allowing for the reduction in momentum from the increase in $\Phi$) can be accounted for primarily by loss from the hot medium.  The flux transfer could in principle be mediated by either mixing or shocks; with  the present simulations, we cannot differentiate between these processes.\footnote{The difference between shock vs. mixing-mediated transfer can be identified in idealized models where the hot gas is injected with a passive scalar; see Schneider et al (2019, in preparation).} In both cases, excess energy flux that is transferred from the hot medium to warm  and intermediate-temperature components in the interaction would be radiated away, as indeed is evident in Figure~\ref{fig:econs}.  

In this section, we use a simple interaction  model to calculate the effective size of warm clouds based on the rate of flux loss from the hot medium.  This allows us to distinguish whether the  hot winds are interacting with many small cloudlets or a few big clouds. 

We adopt a simple model in which (1) the warm gas is in the form of spherical clouds with radius $R_{\rm cl}$ and (2) the hot gas flux is ``absorbed'' by the warm clouds due to interaction.  The flux loss of the hot gas can  then be written as
\begin{eqnarray}\label{eqn:dlnFh}
\deriv{\mathcal{F}_h}{z} = - n_{\rm cl} A_{\rm cl} \mathcal{F}_h,
\end{eqnarray}
where $\mathcal{F}_h$ can either be the momentum or energy flux. Here, $A_{\rm cl}=\pi R_{\rm cl}^2$ is the cross-section of the warm cloud, and $n_{\rm cl}={M_{\rm w}}/(M_{\rm cl}V)$ is the number density of warm clouds,
where $M_{\rm w}=\rho_{\rm w} V_{\rm w}$ is the total mass of warm gas and $M_{\rm cl}=(4/3)\pi R_{\rm cl}^3\rho_{\rm w}$ is the mass of one cloud. 
We can rearrange Equation (\ref{eqn:dlnFh}) as
\begin{eqnarray}\label{eqn:Rcl}
\deriv{ \ln \mathcal{F}_h}{z} = - \frac{3}{4} \frac{f_w}{R_{\rm cl}},
\end{eqnarray}
where $f_w=V_w/V$ is the volume fraction of the warm phase. We note that the interaction (flux loss) is more efficient with a smaller cloud size as the effective cross-section increases with many small clouds.

In Figure~\ref{fig:rcl}(a), we first plot the normalized momentum (solid) and energy (dashed) fluxes, $\mathcal{F}_h/\mathcal{F}_h(z_i)$, in the upper (blue) and lower (red) sides of the disk. We note that the energy and momentum flux losses on a given side are similar to each  other, consistent with the simple physical interaction model.  Since the numerically measured flux is not a smooth function, its numerical derivative (LHS of Equation \ref{eqn:Rcl}) gives very noisy results. For the sake of clarity of presentation, we simply adopt a functional form, $\mathcal{F}_h/\mathcal{F}_h(z_i)=(z/z_i)^{-\alpha}$ with a range of $\alpha$ from 0.5 to 0.9 that describes the general behavior of the flux loss (see the grey shaded region in Figure~\ref{fig:rcl}(a)). With the adopted power-law model, $d\ln \mathcal{F}_h/dz=-\alpha/z$.

Using the volume fraction of the warm phase measured directly from the simulation (Figure~\ref{fig:rcl}(b)), we obtain the cloud size $R_{\rm cl}=(3/4) f_w  z/\alpha$ in Figure~\ref{fig:rcl}(c). Note that the larger $R_{\rm cl}$ corresponds to smaller $\alpha$ and less efficient interaction. The effective value of $R_{\rm cl}$ is generally large, $R_{\rm cl}\gtrsim$ a few 100 pc, implying that at high altitudes the warm phase that interacts with the hot winds exists more likely as one big entity rather than as small cloudlets.  Indeed, the structure seen in Figure~\ref{fig:slicenT} is consistent with this; even though resolved small cloudlets are present, most of the warm medium is in fairly large structures. This results in a smaller effective cross-section for the warm-hot interaction than might be naively expected. 

The interaction between warm clouds and hot, high-velocity flows has been extensively studied to understand whether hot winds can accelerate warm clouds before hydrodynamic instabilities break the clouds apart. Starting from hydrodynamic simulations of adiabatic shocks passing through spherical clouds \citep[e.g.,][]{Klein+94, Xu+95}, shock-cloud interaction simulations have become ever more sophisticated, including e.g., magnetic fields \citep[e.g.,][]{Shin+08, McCourt+15}, radiative cooling \citep[e.g.,][]{Fragile+05,Cooper+09,Scannapieco&Bruggen15}, thermal conduction \citep[e.g.,][]{Bruggen+16}, turbulent clouds \citep[e.g.,][]{Schneider17,Banda-Barragan+19}. Among physical processes considered, radiative cooling seems to prolong the survival of clouds, and, for a sufficiently large cloud size, the mixed gas promotes cooling to grow the total mass of the cool component \citep{Marinacci+10, Armillotta+16, Gronke+18, Gronke+19}.

\citet{Gronke+18} have suggested that above a critical cloud size where $t_{\rm cool,mixed} < t_{\rm cc}$, corresponding to  
\begin{eqnarray}
    R_{\rm cl, crit}& \sim& \frac{v_{\rm wind} t_{\rm cool,mixed}}{\chi^{1/2}}\\
    &=& 109\pc \rbrackets{\frac{v_{\rm wind}}{100\kms}}\rbrackets{\frac{t_{\rm cool,mixed}}{10\Myr}}\rbrackets{\frac{\chi}{100}}^{-1/2},
\end{eqnarray}
clouds will grow in mass.
In the above, we have normalized based on conditions in our simulations at $z\sim1\kpc$, where the density contrast between warm and hot is about $\chi\sim100$, the relative velocity is $v_{\rm wind}\sim 100\kms$, and the cooling time in the intermediate phase is $t_{\rm cool,mix}\sim 1$-10 Myr.  

Here, even though the effective warm cloud size we evaluate is quite large (Figure~\ref{fig:rcl}), we find that there is no significant mass added from the  hot to the warm phase. This apparent discrepancy with the simple critical size estimation can be due to additional complexities that are missing in idealized shock-cloud studies. In  \citet{Gronke+18}, the setup assumes a hot, laminar  winds with a large mass reservoir that is continuously blown, and cooling occurs in a wake behind the cloud. In our simulations, the hot reservoir is limited, and the bursty behavior of outflows leads to intermittent interactions between clouds and winds.  
Indeed, other simulations where the hot wind is turbulent, like our own, have not found evidence for interaction-triggered cooling of the hot medium, and have suggested that lateral turbulent flows limit the downstream condensation (Schneider et al 2019, in preparation).

\section{Summary \& Discussion}\label{sec:summary}

Subsequent to star formation events in galaxies, in many (but not all) cases there will be a successful breakout of a superbubble from the disk of the ISM into circumgalactic space. The eruption drives an outflow consisting of an energetic but low density hot ($T>10^6\Kel$) wind and heavily loaded warm  ($T\sim10^4\Kel$) outflow toward the extra-planar region at $|z| \simgt 1$kpc.  Since feedback quenches individual star formation events by  dispersing dense gas  at the same time as driving outflows, both the star formation and outflows occur in (quasi-periodic) bursts. If the galactic potential is too deep for the moderate-velocity  warm gas to escape, it turns around as an inflow; this is usually called a fountain  flow. Some bursts will be unsuccessful in driving substantial outflows if the outflowing gas is crushed by the return of previously-ejected warm  material.

While hot winds dominate the energy flux and carry substantial momentum and energy into the CGM and galactic halo, in terms of mass the warm outflow/inflow dominates the region within a few kpc surrounding the disk. 
For a complete understanding of extra-planar gas kinematics and dynamics, it is important to investigate how different gas phases interact with each other during inflow/outflow cycles. 

In this paper, we analyze MHD simulations carried out using the TIGRESS numerical framework \citep{Kim&Ostriker17}, targeting star formation, ISM, and galactic potential conditions similar to the solar neighborhood. This simulation is ideal for investigation of galactic outflows with complex interactions because: (1) the simulation duration is quite long (nearly a Gyr), covering many star formation/feedback and outflow/inflow cycles; (2) star formation rates are self-regulated (hence self-consistent with the time-dependent multiphase ISM state) and SNe occur in star clusters and runaways, providing realistic spatio-temporal correlations of feedback sources with  each other and with  the surrounding ISM; and (3) the uniform spatial resolution employed in the simulation is necessary to correctly capture multiphase outflows and interaction between them numerically.
\citet{Kim&Ostriker18} presented an initial analysis of this model to quantify the characteristics of the hot winds and warm fountains. In this paper, we extend the previous analysis to compare with the predictions of a simple ballistic model, and to quantify the exchange of mass, momentum, and energy flux among phases. 

Our main results are summarized as follows:
\begin{enumerate}
    
    \item The predictions of the ballistic model approximate the kinematic distribution of the extra-planar warm gas in the simulation reasonably well at intermediate vertical velocities ($50\kms \lesssim |v_z| \lesssim 100 \kms$). The agreement between model and simulation data worsens at high velocities, where the ballistic model underestimates the amount of warm mass flux almost by a factor 2 (Figure~\ref{fig:ballistic_warm}).  This result indicates that interaction with other phases may partially affect the warm gas kinematics.
   
    \item Cooling of gas at intermediate temperatures transfers some mass flux from the intermediate phase to the warm phase. From $|z|=1$~kpc to $|z|=3$~kpc, the warm gas has gained almost $50-100\%$ of its initial mass flux from the intermediate phase (Figure~\ref{fig:mfluxterms}, \ref{fig:mcons}). However, 
    even when the v-PDFs of warm and intermediate gas are summed, a discrepancy between the ballistic model and the simulation results persists (Figure~\ref{fig:ballistic_warm_int}).  The discrepancy is particularly large on the lower side of the disk.
    
    \item The missing piece is the exchange of momentum flux between warm and hot gas. The amount of momentum flux transferred from the hot to the warm phase within $|z|=3$~kpc is $50\%$ of the momentum flux of the warm phase at the launching height, $z=1$~kpc (Figure~\ref{fig:pcons}). The warm gas gains considerable momentum flux from the hot gas, but negligible mass flux. This results in acceleration  of the warm gas. 
    
    \item The energy flux of the hot phase shows huge losses between $|z|=1$-3~kpc, but no other phase increases its energy flux. The energy flux lost from the hot gas is $\sim 20-30$ times the initial energy flux of the warm phase. The loss of energy flux from the hot medium (including the direct energy injection from SNe exploding in extra-planar regions) without a corresponding energy flux gain in another phase  can  be explained by strong cooling in both warm and intermediate phases.
    
    \item Based on the flux changes in the hot phase and a simple interaction model with spherical clouds, we derive the effective cloud size of a few 100 pc (Figure~\ref{fig:rcl}). The hot phase loses momentum and energy through interaction with a few large warm clouds rather than many small cloudlets.
\end{enumerate}

We emphasize that a number of features particular to our simulation has enabled the  quantitative analysis of this paper. First, the simulation reaches  a quasi-steady state in an average sense. Vertical dynamical equilibrium holds in the sense that the time-averaged momentum flux difference balances the weight of the gas (Figure~\ref{fig:vertical_equilibrium}a), while bursty star formation and inflow/outflow cycles are evident from the horizontally averaged space-time diagram of mass fluxes (Figure~\ref{fig:mflux}).  Covering several \emph{self-consistent} star formation/feedback cycles ($\sim$ a few hundred Myr for solar neighborhood conditions) is important for studying fountains in detail; other local simulations have had much shorter durations \citep[e.g.,][]{Gatto+17,Kannan+18}. 

As part of our analysis, we separate gas into three different thermal phases and make use of horizontally and temporally averaged vertical profiles of physical quantities. Uniform resolution both near the midplane and in the extra-planar regions allows a fair phase separation within the simulated volume. In simulations with adaptive resolutions (using semi-Lagrangian or adaptive mesh refinement code), typically the resolution is progressively poorer as flows move outward.  Without sufficient resolution and phase separation at all altitudes, investigation of interaction between phases in outflows is challenging, especially in the extra-planar region \citep[e.g.,][]{Muratov+15, Kannan+18,Hu19}.

To assess the extent of interactions 
we investigate the exchanges of the mass, momentum, and energy fluxes between thermal phases using the profiles of flux differences between the height of interest and launching (Figures~\ref{fig:mcons},\ref{fig:pcons},\ref{fig:econs}). To isolate the exchange of fluxes between phases, it is important to factor out all other potential flux changes, including (1) the net mass change in the warm phase by the imperfectness of the steady-state assumption, (2) the momentum loss as the gas climbs up the gravitational potential, and (3) energy loss by cooling and gain by direct SN explosion. After taking these additional ``source'' terms into account, we can directly link the excess/deficit of flux in one phase to the deficit/excess of flux in another phase.

A possible caveat that should  be kept in mind is that the simulations analyzed here did not include thermal conduction, which might
alter the interaction between different gas phases. The effect of thermal conduction is to transfer thermal energy from a hotter to a colder medium, potentially leading warm gas to evaporate in the surrounding hotter material \citep[e.g.,][]{Weaver+77,Cowie+77,Bruggen+16,Armillotta+17}. However, whether an effective phase transition -- with resulting addition of mass to the intermediate/hot phase -- occurs depends on the balance between thermal conduction and radiative cooling. 
If the cooling time is smaller than the evaporation time, the newly-generated intermediate gas will cool quickly, returning mass to the warm phase and radiating away the thermal energy initially transferred from the hot to the warm phase. In addition, interfaces where conduction occurs are likely  also to have turbulent mixing.  Even if conductively-heated gas does not have a short cooling time, mixing with nearby dense gas may lead to efficient radiation of the energy conducted out of the hot medium.  
In general, it can be expected that conductive heating increases the mass of hot gas and therefore potential mass loading of hot winds, while mixing that leads to cooling limits the energy loading of winds 
\citep[e.g.,][]{ElBadry19}. 

We note however that the effect of thermal conduction might be mitigated by the presence of magnetic fields.  
The motion of the conducting electrons is parallel to the magnetic field lines, so that the efficiency of thermal conduction is strongly reduced in the transverse direction \citep[e.g.,][]{Braginskii65,Orlando+08}.

The results presented in this work are highly relevant to
interpretation of extra-planar gas kinematics in Milky Way-like galaxies. Recently, there have been some attempts to model \hi\ and H$\alpha$ extra-planar gas in the Milky-Way and nearby star-forming galaxies by using ballistic models \citep{Collins+02, Fraternali&Binney06, Marasco+12}. Although these models have been able to reproduce the extra-planar gas profiles, they are unable to reproduce its kinematics. Interaction with surrounding gas is required to explain the observations \citep{Fraternali&Binney08, Marasco+12}, although the origin of this interaction has been poorly investigated.

At this stage, we cannot generalize our results to galactic environments that differ largely from the Milky Way's Solar neighborhood. Galaxy properties affect the gas properties at the launch location, which, in turn, drive the level of interaction during the outflow evolution.
In the simulation studied here,  for example, the mass and momentum fluxes in the warm medium are larger than other phases at launch, suggesting at lowest order that a ballistic fountain model may be appropriate for this low-filling-factor component, neglecting interactions with other components.
In simulations  of dwarf galaxies, 
\citet{Hu19} instead finds that the interaction between hot and warm gas is very effective in accelerating warm gas to velocities larger than the galaxy escape velocity. This might be due to the combined effect of a weaker gravitational potential (lower galaxy escape velocity) and higher energy loading factors of the hot gas. An important next step for the present work will be to apply our framework to simulations performed with conditions representative of different galactic environments.

Finally, we highlight that thorough study of the outflow properties in high-resolution local disk simulations, such as the one performed in this paper, can provide the detailed information required to build sub-grid models for wind launching in cosmological galaxy formation simulations.

\acknowledgements

AV received travel support from ITS, SERB, Government of India and would like to thank Biman B. Nath and Prateek Sharma for useful discussions and encouragement. The work of E.~C.~O. was supported  by NASA ATP grant NNX17AG26G, and C.-G.~K. was supported by grant 528307 from  the Simons Foundation. L.~A. acknowledges support from the Australian Research Council's Discovery Projects and Future Fellowships funding schemes, awards DP190101258 and FT180100375. This work was initiated as a project for the Kavli Summer Program in Astrophysics held at the Center for Computational Astrophysics of the Flatiron Institute in 2018. The program was co-funded by the Kavli Foundation and the Simons Foundation, and we thank them for their generous support.  Resources supporting this work were provided in part by the NASA High-End Computing (HEC) Program through the NASA Advanced Supercomputing (NAS) Division at Ames Research Center, in part by the Princeton Institute for Computational Science and Engineering (PICSciE) and the Office of Information Technology’s High Performance Computing Center, and in part by the National Energy Research Scientific Computing Center, which is supported by the Office of Science of the U.S. Department of Energy under Contract No. DE-AC02-05CH11231.


\end{document}